\documentclass[a4paper,fleqn,usenatbib]{mnras}

\usepackage[T1]{fontenc}
\usepackage{ae,aecompl}

\usepackage{graphicx}	
\usepackage{amsmath}	
\usepackage{amssymb}	
\usepackage{natbib}
\usepackage{nccmath}
\usepackage{epsfig}
\usepackage{epstopdf}
\usepackage{balance}
\usepackage{mathrsfs}

\title[Eccentricity distributions of EBBHs in GNs]{Eccentricity distributions of eccentric binary black holes in galactic nuclei}

\author[J. Tak\'{a}tsy, B. B\'{e}csy, and P. Raffai]{
J. Tak\'{a}tsy$^{1}$\thanks{E-mail: \href{mailto:takatsyj@caesar.elte.hu}{takatsyj@caesar.elte.hu} (JT)},
B. B\'{e}csy$^{1}$,
P. Raffai$^{1,2}$
\\
$^{1}$Institute of Physics, ELTE E\"{o}tv\"{o}s University, 1117 Budapest, Hungary\\
$^{2}$MTA-ELTE Extragalactic Astrophysics Research Group, 1117 Budapest, Hungary
}

\date{Accepted XXX. Received YYY; in original form ZZZ}

\pubyear{2018}

\begin{document}
\label{firstpage}
\pagerange{\pageref{firstpage}--\pageref{lastpage}}
\maketitle

\begin{abstract}
Galactic nuclei are expected to be one of the main sites for formations of eccentric binary black holes (EBBHs), with an estimated detection rate of \mbox{$\mathcal{O}(1-100$ yr$^{-1})$} with Advanced LIGO (aLIGO) detectors operating at design sensitivity. The two main formation channels of these binaries are gravitational capture and the secular Kozai-Lidov mechanism, with expectedly commensurable formation rates. We used Monte Carlo simulations to construct the eccentricity distributions of EBBHs formed through these channels in galactic nuclei, at the time their gravitational-wave signals enter the aLIGO band at $10$~Hz. We have found that the proportion of binary black holes entering the aLIGO band with eccentricities larger than $0.1$ is $\sim 10$ percent for the secular Kozai-Lidov mechanism, and $\sim 75$ percent for gravitational capture. We show that if future EBBH detection rates with aLIGO will be dominated by EBBHs formed in galactic nuclei, then the proportions of EBBHs formed through the two main channels can be constrained to a $\Delta \mathcal{F}= 0.2$ wide one-sigma confidence interval with a few tens of observations, even if parameter estimation errors are taken into account at realistic levels.
\end{abstract}

\begin{keywords}
black hole physics -- gravitational waves -- galaxies: nuclei
\end{keywords}



\section{Introduction}
\label{sec:Introduction}
\parskip 0pt

Second generation gravitational-wave (GW) detectors are state-of-the-art, ground-based, L-shaped interferometers with kilometer-scale arms. The network of Advanced LIGO (aLIGO) GW detectors \citep{aasi2015b}, consisting of aLIGO-Hanford and aLIGO-Livingston, finished its first observing run (O1) in January 2016. During O1, this network achieved the first detections of GWs from coalescences of binary black holes (BHs) \citep{abbott2016}. The two aLIGO detectors carried out their second observing run between November 2016 and August 2017, with an improved network sensitivity compared to O1, achieving further observations of binary BHs \citep{abbott2017b,abbott2017d} and the first observation of a binary neutron star \citep{abbott2017c}. Second generation GW detectors also include Advanced Virgo \citep{acernese2015}, which joined the aLIGO observing run in August 2017. Two additional GW detectors are planned to join aLIGO and Advanced Virgo: the Japanese KAGRA is under construction and is planned to begin its first observing run in 2020, while LIGO-India is expected to become operational in 2024 at the earliest \citep{abbott2018}.

The most common sources for GW detections with second generation detectors are binaries of stellar-mass BHs \citep{abadie2010}. These systems can form through binary stellar evolution \citep[e.g.][]{postnov2014}, as well as through dynamical processes \citep[e.g.][]{park2017}. Dynamical formation involves non-dissipative processes, such as strong gravitational encounters and dynamical three-body formation, but it also includes gravitational capture, as a dissipative process. Binary BHs formed dynamically in dense stellar environments generally exhibit higher eccentricities compared to those forming in isolated environments in galactic fields \citep[see e.g.][]{breivik2016}. Since GW emission tends to circularize the orbit of binary BHs, most of the binaries are expected to possess negligible eccentricities (usually $e \lesssim 10^{-3}$) at the time their GW frequency reaches \mbox{$10$ Hz}, i.e. the nominal low-frequency limit of aLIGO detectors \citep{breivik2016, rodriguez2016}. Several studies have shown that signal search methods using circular binary templates \citep[see e.g.][and references therein]{capano2016} are unsuitable for finding binary BHs with $e>0.1$ when entering the aLIGO band \citep{brown2010,huerta2013}. Some processes, including the ones we studied here in details, are predicted to produce such $e>0.1$ binaries. Hence within the scope of our work, we define eccentric binary BHs (EBBHs) as binary BHs having eccentricities $e>0.1$ at the time their GW frequency reaches $10$ Hz.

According to theoretical models, we expect most EBBHs to form in dense stellar clusters, such as galactic nuclei (GNs) and globular clusters (GCs) \citep{portegies2000, kocsis2006, oleary2009}. Two of the main formation channels of these binaries are the following: (i) Initially unbound BHs can form binaries during close encounters, if they lose enough kinetic energy to become bound. In almost all cases, this energy loss occurs through GW emission, possibly resulting with an EBBH \citep[e.g.][]{peters1963, oleary2009}. (ii) EBBHs can also form in hierarchical triples through the secular Kozai-Lidov mechanism \citep[e.g.][]{wen2003, antoniniperets2012, antonini2014, antognini2016, hoang2018}. During this process, binary BHs are expected to reach very high eccentricities \citep[with $1-e$ ranging from $10^{-2}$ to $10^{-5}$, similarly to EBBHs formed through gravitational capture, see][]{gondan2017} with a merger rate comparable with that of EBBHs formed through gravitational capture.

In GNs the expected merger rate density of EBBHs is \mbox{$\sim 0.8$ Gpc$^{-3}$yr$^{-1}$} for EBBHs formed through gravitational capture \citep{oleary2009}. \cite{hoang2018} showed that the total merger rate density of all binary BHs formed through the secular Kozai-Lidov mechanism in GNs is \mbox{$\sim 1-3$ Gpc$^{-3}$yr$^{-1}$}, which, based on considerations derived from our results described in Section \ref{sec:KozaiGN}, corresponds to a merger rate density of \mbox{$\mathcal{O}(0.1$ Gpc$^{-3}$yr$^{-1}$)} for EBBHs. \cite{samsing2018} and \cite{rodriguez2018} recently showed that the formation rate of EBBHs produced by binary-single interactions in GCs may be comparable to the formerly mentioned formation rates with an expected merger rate density of \mbox{$0.05-0.5$ Gpc$^{-3}$yr$^{-1}$}. Another alternative for EBBH formation in GNs and GCs is through multi-body interactions \citep[e.g.][]{antoninirasio2016,zevin2018}, although the merger rate from this channel is expected to be at least an order of magnitude below the rates for the previously mentioned channels.

Binary BHs formed through gravitational capture are expected to be detectable up to a few Gpc by aLIGO detectors operating at design sensitivity, with estimated total detection rates of \mbox{$\mathcal{O}(1-100$ yr$^{-1}$)} for GNs and $\mathcal{O}$(1 yr$^{-1}$) for GCs \citep[see e.g.][]{kocsis2006, oleary2009}. Although the exact ratio of detection rates for the different EBBH formation channels is yet unknown, we also expect a commensurable detection rate for binary BHs formed through the secular Kozai-Lidov mechanism \citep[see e.g.][]{antonini2014,hoang2018}. Note that these detection rates are non-negligible compared to the rates predicted for circular binary BHs. In fact, the reconstructed spin configuration of one of the observed binary BHs \citep[GW170104, see][]{abbott2017a} suggests that it may have formed dynamically in a dense stellar cluster. The relatively high masses of the observed binary BHs are also consistent with the assumption that these binaries were formed through dynamical processes in such environments \citep[see e.g.][]{mckernan2017}.

According to the leading order results in \cite{peters1963}, compared to circular binaries, EBBHs are detectable from greater distances and in a wider mass range due to the strength and spectral richness of their GW signals \citep{oleary2009, kocsislevin2012}. By detecting these signals, in principle, many parameters of their sources can be reconstructed, including their orbital parameters (eccentricity and pericenter distance) at the time the frequency of the binary's GW signal reaches $10$ Hz. Numerical tools for simulating GW signals of EBBHs already exist \citep[see e.g.][]{levin2011, csizmadia2012, east2013} and are available for testing search algorithms. There are also search methods available aiming to find GW signals of highly eccentric EBBHs \citep{tai2014,tiwari2016} and EBBHs with $e<0.6$ \citep{coughlin2015}. For reconstructing circular binary parameters, several algorithms are used, e.g. BAYESTAR \citep{singer2016}, LALInference \citep{veitch2015}, and GstLAL \citep{cannon2012, privitera2014}. Coherent WaveBurst \citep{klimenko2016} and BayesWave \citep{cornish2015,becsy2017,millhouse2018} algorithms offer ways for model-independent parameter reconstruction. The development of algorithms aiming to recover EBBH parameters is still underway.

In the work presented in this paper we used Monte Carlo simulations to construct the eccentricity distributions of EBBHs formed through gravitational capture and the secular Kozai-Lidov mechanism in GNs, at the time their GW frequency enters the aLIGO band. We applied a set of numerical simulations introduced in \cite{gondan2017} to determine the expected parameter distributions of EBBHs formed through gravitational capture, however we extended them by introducing randomized masses for the central supermassive BHs (SMBHs). For EBBHs formed through the secular Kozai-Lidov mechanism, we solved the octupole-order equations of motion of \cite{naoz2013} in which we included gravitational radiation, while making reasonable assumptions on the initial parameter distributions of triple systems. In this paper we also show that if future EBBH detection rates with aLIGO will be dominated by EBBHs formed in GNs, then it is feasible to constrain the proportions of EBBHs formed through the two main channels (i.e. the branching ratios). We performed a statistical test to determine the minimum number of EBBH observations needed to constrain the branching ratio of gravitational capture (from now on, denoted by $\mathcal{F}\mathrm{_{GC}}$) to an arbitrarily chosen $\Delta \mathcal{F}\mathrm{_{GC}} = 0.2$ wide one-sigma confidence interval (note that for symmetrical errors $\Delta \mathcal{F}\mathrm{_{GC}} = 0.2$ corresponds to a $\pm10$ percent range of the total $[0,1]$ interval). First, we carried out our tests with the idealized case of having no errors on the estimated parameters. We then repeated our analysis, assuming statistical errors on the estimated parameters. Since EBBH parameter estimation methods are still in development, we only considered a set of plausible errors consistent with the results of previous studies \citep[see][]{gondan2018,lower2018,gondankocsis2018}.

This paper is organized as follows. In Section \ref{sec:EBBH} we review the mechanism of EBBH formation through gravitational capture, as well as through the secular Kozai-Lidov mechanism, and describe the evolution of EBBH orbits using the leading order results of \cite{peters1964}. In Section \ref{sec:GN} we introduce the model we used for GNs in our work and describe the numerical methods we applied in our Monte Carlo simulations to generate EBBHs. We present our results in Section \ref{sec:KozaiGN}. Finally, we offer our conclusions and summarize the implications of our work in Section \ref{sec:Conclusion}.

\section{Eccentric Binary Black Holes}
\label{sec:EBBH}

In this section we summarize the formational mechanism of EBBHs through gravitational capture (see Section \ref{ssec:InPar}) and through the secular Kozai-Lidov mechanism (see Section \ref{ssec:SecKozai}). First, we introduce the equations to calculate the initial orbital parameters for EBBHs formed through gravitational capture, as well as their evolution during the eccentric inspiraling of the two BHs. We then consider the possible disruptions of formed binaries by gravitational interactions with third objects. Finally, we describe the Kozai-Lidov mechanism as an alternative formational process of EBBHs, and show the importance of using equations of motion beyond the quadrupole approximation.

In our calculations, we use the geometric unit system ($G=c=1$). We denote the total mass of the two BHs forming a binary by \mbox{$M=m_1+m_2$}, and their symmetrical mass ratio by \mbox{$\eta=m_1 m_2 /M^2$}, where $m_1$ and $m_2$ are the masses of the BHs, defined as $m_1 \geq m_2$. The reduced mass of the binary is \mbox{$\mu=M \eta$}. We describe the orbital evolution in a quasiperiodic manner. We define the dimensionless pericenter distance as $\rho_{\mathrm{p}} = r_{\mathrm{p}}/M$, where $r_{\mathrm{p}}$ is the pericenter distance. The mean orbital frequency from Kepler's law is \mbox{$f_{\mathrm{orb}} = (2\upi M)^{-1} \rho_{\mathrm{p}}^{-3/2} (1-e)^{3/2}$}, where $e$ is the eccentricity of the orbit. Note that in this quasiperiodic treatment, the time evolution of the system is incorporated in the time-dependence of the orbital parameters $e$ and $\rho_{\mathrm{p}}$. The following discussion is based on \cite{oleary2009} and \cite{gondan2017}.

\subsection{Formation through gravitational capture}
\label{ssec:InPar}

During a close encounter, two BHs can form a binary if they emit enough energy in the form of GWs to become bound. Due to the relativistic nature of such events and the fact that the velocity dispersion in GNs is $\sigma \ll 1$, the encounters are always nearly parabolic \citep{quinlan1987,lee1993}, and the binaries formed this way are always highly eccentric \citep[$1-e_0^2\sim10^{-1}-10^{-5}$, where $e_0$ is the initial eccentricity of the formed binary, see e.g.][]{gondan2017}. The change in energy and angular momentum due to GW emission during the encounter is

\begin{ceqn}
\begin{equation}
\delta E = - \frac{85\pi \eta^2 M^{9/2}}{12 \sqrt{2} r_{\mathrm{p0}}^{7/2}} ,
\end{equation}
\end{ceqn}

\begin{ceqn}
\begin{equation}
\delta L = - \frac{6\pi \eta^2 M^4}{r_{\mathrm{p0}}^2} ,
\label{eq:deltaL}
\end{equation}
\end{ceqn}
where

\begin{ceqn}
\begin{equation}
r_{\mathrm{p0}}=\left( \sqrt{\frac{1}{b^2} + \frac{M^2}{b^4 w^4}}+\frac{M}{b^2 w^2} \right)^{-1} ,
\end{equation}
\end{ceqn}
we denoted the initial values of orbital parameters with a '0' symbol in the lower index, $b$ is the impact parameter of the encounter, and $w$ is the relative velocity of the two BHs \citep[see][]{peters1963,turner1977}.

The initial orbital parameters of the binary are determined by the final energy ($E_{\mathrm{fin}}$) and final angular momentum ($L_{\mathrm{fin}}$) of the system after the first encounter. \cite{oleary2009} showed that $\delta L$ (see Eq. \ref{eq:deltaL}) is negligible for nearly all first encounters, therefore, for simplicity, we set $\delta L = 0$. In case the BHs emit enough energy to have $E_{\mathrm{fin}}<0$, the system becomes bound after the first encounter, with an initial semi-major axis

\begin{ceqn}
\begin{equation}
a_0 = - \frac{\eta M^2}{2 E_{\mathrm{fin}}} ,
\end{equation}
\end{ceqn}
and initial eccentricity

\begin{ceqn}
\begin{equation}
e_0 = \sqrt{1+2\frac{E_{\mathrm{fin}}b^2w^2}{M^3 \eta}} .
\end{equation}
\end{ceqn}

Note that there is an upper and lower limit for the $b$ impact parameter of encounters that produce EBBHs for fixed values of $w$. The upper limit is set by the condition $E_{\mathrm{fin}}<0$, since a distant encounter of BHs does not produce bound binary system. Additionally, there is a lower limit set by the fact that we require BHs to avoid direct collision during their first encounter \citep{kocsis2006,oleary2009}. Therefore, in leading order, $b$ must satisfy,

\begin{ceqn}
\begin{equation}
b_{\mathrm{min}} \equiv \frac{4 M}{w} <b<\left(\frac{340\pi \eta}{3} \right)^{1/7} \frac{M}{w^{9/7}} \equiv b_{\mathrm{max}} .
\label{eq:bminmax}
\end{equation}
\end{ceqn}

The leading order approximation in the allowed range of impact parameters is in excellent agreement with $2.5$ and $3.5$ order post-Newtonian simulations \citep{kocsislevin2012}. Note that in realistic GNs, relativistic corrections only modify the capture cross section by less than $10$ percent \citep{kocsis2006,oleary2009}.

\subsection{Evolution of isolated eccentric binaries}
\label{ssec:OrbEv}

The time derivative of the eccentricity in the leading order approximation is given by the equation \citep{peters1964}

\begin{ceqn}
\begin{equation}
\frac{\mathrm{d}e}{\mathrm{d}t} = - \frac{304 \eta e(1-e)^{3/2}}{15 M \rho_{\mathrm{p}}^4 (1+e)^{5/2}} \left( 1+ \frac{121}{304} e^2 \right) .
\label{eq:eevolution}
\end{equation}
\end{ceqn}
By utilizing the formula for d$a/$d$t$ \citep[see Eq. 5.6 in][]{peters1964}, the time derivative of $\rho_{\mathrm{p}}$ can be expressed as

\begin{ceqn}
\begin{align}
\frac{\mathrm{d}\rho_{\mathrm{p}}}{\mathrm{d}t} = - \frac{64 \eta e(1-e)^{1/2}}{5 M \rho_{\mathrm{p}}^3 (1+e)^{7/2}} \left( 1+ \frac{73}{24} e^2 + \frac{37}{96} e^4 \right) + \nonumber \\
+ \frac{304 \eta e(1-e)^{3/2}}{15 M \rho_{\mathrm{p}}^3 (1+e)^{5/2}} \left( 1+ \frac{121}{304} e^2 \right) .
\label{eq:rhoevolution}
\end{align}
\end{ceqn}
Dividing Eq. (\ref{eq:rhoevolution}) by Eq. (\ref{eq:eevolution}) yields a separable differential equation for $\rho_{\mathrm{p}}(e)$, which gives the solution \citep{peters1964}:

\begin{ceqn}
\begin{equation}
\rho_{\mathrm{p}} = \frac{c_0 e^{12/19}}{(1+e)M} \left( 1+\frac{121}{304}e^2 \right)^{\frac{870}{2299}} ,
\label{eq:rhoe}
\end{equation}
\end{ceqn}
where $c_0$ is a function of the initial parameters that can be determined by setting $e=e_0$ and $\rho_{\mathrm{p}}=\rho_{\mathrm{p0}}$ in Eq. (\ref{eq:rhoe}).

The orbital equations Eq. (\ref{eq:eevolution}-\ref{eq:rhoe}) are only valid for \mbox{$\rho_{\mathrm{p}} \gg 2$}, when the BHs are relatively far from each other. At a certain point the binary reaches its last stable orbit (LSO), after which the evolution is no longer quasi-periodic, and the BHs in the binary quickly merge. In the leading order approximation for zero spins and a finite mass ratio, $e_{\mathrm{LSO}}$ can be given by solving the following equation numerically \citep{cutler1994}:

\begin{ceqn}
\begin{equation}
\rho_{\mathrm{p}}(e_{\mathrm{LSO}}) = \frac{6+2e_{\mathrm{LSO}}}{1+e_{\mathrm{LSO}}} ,
\end{equation}
\end{ceqn}
where $\rho_{\mathrm{p}}(e_{\mathrm{LSO}})$ is obtained by substituting $e_{\mathrm{LSO}}$ to \mbox{Eq. (\ref{eq:rhoe})}. Unlike GW signals of circular binaries, the ones from EBBHs cover a broad band of frequencies, where the maximum power occurs at a harmonic $n_m \gg 1$ for $e \approx 1$ ($f_m=n_m f_{\mathrm{orb}}$ is the frequency of the GW signal's $m^{\mathrm{th}}$ harmonic in Fourier space). \cite{wen2003} showed that an approximate expression can be derived for this "peak harmonic" in the form of

\begin{ceqn}
\begin{equation}
n_{\mathrm{peak}} \approx \frac{2(1+e)^{1.1954}}{(1-e^2)^{3/2}} .
\end{equation}
\end{ceqn}
In this context, GWs of EBBHs are often considered entering the aLIGO band when the GW frequency,

\begin{ceqn}
\begin{equation}
f_{\mathrm{GW}} = n_{\mathrm{peak}} \frac{M^{1/2}}{2\pi a^{3/2}} ,
\label{eq:fgw}
\end{equation}
\end{ceqn}
reaches $10 \: \mathrm{Hz}$ \citep[see e.g.][and references therein]{wen2003,gondan2017}, where both \mbox{$n_{\mathrm{peak}}=n_{\mathrm{peak}}(e)$} and \mbox{$a=a(e)$} are functions of $e$. Hence, we also use this criterion in our study.

By substituting Eq. (\ref{eq:rhoe}) to Eq. (\ref{eq:eevolution}), the merger time can be expressed as an integral over eccentricities from $e_{\mathrm{LSO}}$ to $e_0$ \citep{gondan2017}:

\begin{ceqn}
\begin{equation}
t_{\mathrm{merge}}(m_1,m_2,r_{\mathrm{p0}},e_0) = \frac{15 c_0^4}{304 M^3 \eta} \mathcal{I}(e_{\mathrm{LSO}},e_0) ,
\end{equation}
\end{ceqn}
where

\begin{ceqn}
\begin{equation}
\mathcal{I}(e_{\mathrm{LSO}},e_0) = \int\limits^{e_0}_{e_{\mathrm{LSO}}} \frac{e^{29/19}(1+\frac{121}{304}e^2)^{\frac{1181}{2299}}}{(1-e^2)^{3/2}} \mathrm{d}e .
\end{equation}
\end{ceqn}

GNs are dense stellar environments, therefore there is a possibility that a third object interacts with the binary during the inspiral. Binary-single interactions can be divided into three major categories: weak perturbations, strong perturbations, and close interactions \citep{samsing2014}. Weak and strong perturbations ultimately lead to the merger of the binary, while close encounters can lead to several different outcomes. According to \cite{oleary2009}, the characteristic timescale of a binary-single interaction can be approximated as

\begin{ceqn}
\begin{equation}
t_{\mathrm{enc}} \approx \frac{w^3}{12\pi M^2 n_{\mathrm{tot}}} ,
\end{equation}
\end{ceqn}
where $n_{\mathrm{tot}}=n_{\mathrm{tot}}(r)$ is the combined number density of all objects residing in a GN (we assume that GNs are spherically symmetric, and $r$ denotes the radial distance from their center). Similarly to \cite{gondan2017}, we assume that binaries satisfying the condition $t_{\mathrm{enc}}>t_{\mathrm{merge}}$ avoid disruption by a third object, and those that do not are disrupted. Note that this assumption overestimates the number of those binary-single interactions that lead to the disruption of the binary. However, \cite{gondan2017} showed that even this assumption results in a negligible number of disrupted binaries.

\subsection{The secular Kozai-Lidov mechanism}
\label{ssec:SecKozai}

Triple systems are also believed to play an important role in the formation of EBBHs \citep[see e.g.][]{millerhamilton2002, thompson2011, wen2003} through secular evolution (i.e. coherent perturbations on timescales much longer than the orbital period), and specifically through the so-called secular Kozai-Lidov mechanism \citep{kozai1962,lidov1962}. Inside a GN, where the dynamics is determined by a central SMBH, binary BHs automatically form a triple system with the SMBH. Previous studies suggested a considerable aLIGO detection rate of \mbox{$\mathcal{O}$(1 yr$^{-1}$)} for mergers of EBBHs formed through the Kozai-Lidov mechanism \citep[see e.g.][]{antonini2014,hoang2018}. \cite{wen2003} calculated the eccentricity distribution for such binaries merging inside GCs using orbit-averaged approximations, \cite{antonini2014} however showed the breakdown of these assumptions for triples with high inclination and with an outer BH at a moderate distance. \cite{antonini2014} and \cite{fragione2018} also showed that compared to the secular approach, direct N-body simulations predict higher eccentricities for EBBHs entering the aLIGO band. Consistently with other studies \citep[e.g.][]{antoniniperets2012,vanlandingham2016}, they found that a fraction of these binaries can have $e>0.1$ when they reach the aLIGO band. However, they did not study the formation of such hierarchical triples, but instead, they made assumptions for the distributions of initial parameters.

A triple system consists of an inner binary with masses $m_1$ and $m_2$ (defined as $m_1 \geq m_2$), and a third object with mass $m_3$ (in our case the third object is always the central SMBH). It is convenient to describe the orbit using Jacobi coordinates \citep[see e.g.][p. 441-443]{murray2000}. In this treatment, the dominant motion of the triple system can be divided into two separate Keplerian orbits: one is the motion of objects 1 and 2 relative to each other, and the other one is the motion of object 3 relative to the center of mass of the inner binary. We denote the relative position vectors, semi-major axes, and eccentricities of these systems by $\mathbf{r}_1$ and $\mathbf{r}_2$, $a_1$ and $a_2$, and $e_1$ and $e_2$, respectively. If the third object is sufficiently distant from the inner binary, a perturbative approach can be used to desribe the evolution of the system. In the usual secular approximation \citep[e.g.][]{marchal1990}, the two orbits exchange angular momentum, but not energy. Hence, the eccentricities and orientations of the orbits can change, but the semi-major axes cannot. The coupling term in the Hamiltonian can be written in a power series expansion of parameter $\alpha = a_1/a_2$ \citep[see e.g.][]{harrington1968}, where the Hamiltonian is

\begin{equation*}
\mathcal{H} = \frac{m_1 m_2}{2a_1} + \frac{(m_1+m_2)m_3}{2a_2} +
\end{equation*}
\vspace{-0.5cm}
\begin{ceqn}
\begin{equation}
+ \frac{1}{a_2} \sum\limits_{j=2}^{\infty} \alpha^j M_j \left( \frac{r_1}{a_1} \right)^j \left( \frac{a_2}{r_2} \right)^{j+1} P_j(\mathrm{cos}\Phi) ,
\end{equation}
\end{ceqn}
$P_j$ is the $j$th-degree Legendre polynomial, $\Phi$ is the angle between $\mathbf{r}_1$ and $\mathbf{r}_2$, and

\begin{ceqn}
\begin{equation}
M_j = m_1 m_2 m_3 \frac{m_1^{j-1}+(-1)^j(m_2)^{j-1}}{(m_1+m_2)^j} .
\end{equation}
\end{ceqn}

When analyzing these hierarchical triple systems, it is particularly convenient to adopt the canonical coordinates known as Delaunay's elements \citep[e.g.][]{valtonen2006}. The coordinates are the mean anomalies, $l_1$ and $l_2$, the longitudes of ascending nodes, $h_1$ and $h_2$, and the arguments of periastron, $g_1$ and $g_2$. One can then eliminate the short-period terms by a canonical transformation to get equations that describe the long-term evolution of the triple system. This technique is known as the \textit{Von Zeipel transformation} \citep[see e.g.][]{brouwer1959}. \cite{naoz2013} showed that going beyond the quadrupole-order approximation is necessary in order to acquire highly eccentric Kozai-excitations. They also pointed out a common mistake in the Hamiltonian treatment of these hierarchical systems in previous studies, which could lead to erroneous conclusions in the test particle limit. The full octupole-order equations of motion can be found in \cite{naoz2013}.

\begin{figure}
	\includegraphics[width=\columnwidth]{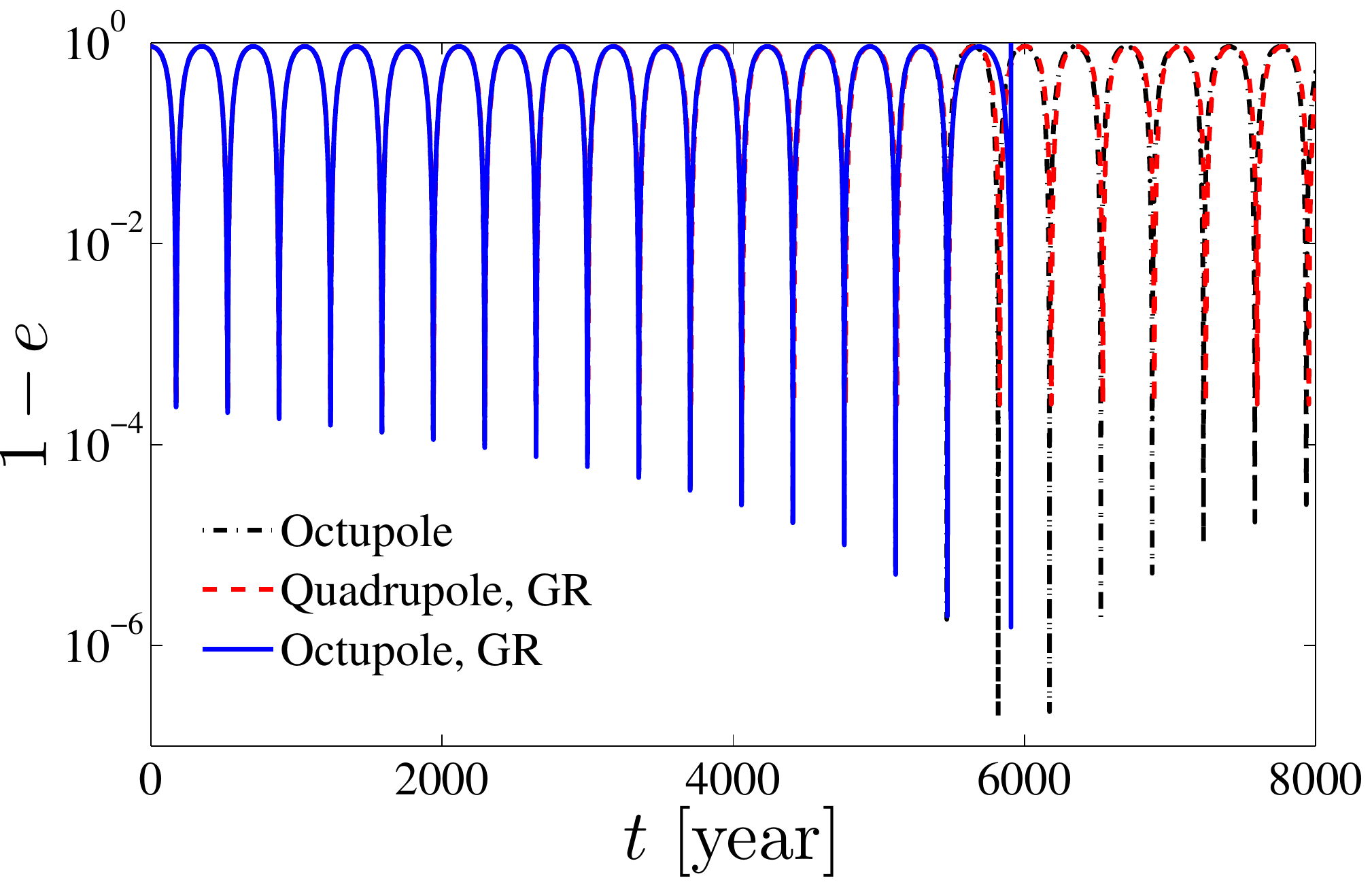}
    \caption{An example for the evolution of eccentricity $e$ in a typical hierarchical triple system, where the general relativistic terms become significant after a few Kozai-cycles. The inner binary contains BHs with masses $m_1=16.9\:M_{\odot}$ and $m_2=6.5\:M_{\odot}$, the outer BH has a mass $m_3=4.4 \times 10^6\:M_{\odot}$. The inner orbit has $a_1=16.7$ AU, while the outer orbit has $a_2=3500$ AU. The initial eccentricities are $e_1=0.078$ and $e_2=0.37$, and the initial relative inclination is $I=89^{\circ}$. Additionally, the initial arguments of periastron are $g_1=1.2^{\circ}$ and $g_2=105^{\circ}$. Using the quadrupole-order equations of motion, including gravitational radiation does not lead to a merger of the inner binary (dashed red line). An expansion beyond this approximation without gravitational radiation (dash-dotted black line) leads to a chaotic behaviour, which, including gravitational radiation, leads to an eventual merger of the inner binary (solid blue line).}
    \label{fig:KozaiQPOP}
\end{figure}

We utilized a fourth-order Runge-Kutta method with adaptive stepsize control to solve the octupole-order equations of motion of \cite{naoz2013}, including terms from gravitational radiation effects. Fig. \ref{fig:KozaiQPOP} shows the eccentricity evolution of the inner binary in a typical hierarchical triple system, where the general relativistic terms become dominant after a few Kozai-cycles. Using the quadrupole-order approximation results in closed, periodic orbits in the phase space of parameters. However, by expanding our equations to the octupole-order, the behaviour of triple systems becomes chaotic. This can lead to extremely high eccentricities for the inner binaries, in which case the gravitational radiation term becomes dominant. As a result, the inner binary ultimately merges. Note that going beyond the octupole-order expansion is unnecessary in most cases, since this approximation already exhibits the system's chaotic nature, and all higher-order terms take effect on a longer timescale. In addition, yet another drastic change in the qualitative behaviour of the system is not expected, since all the remaining integrals of motion beyond the octupole-expansion are the total energy and the components of the total angular momentum, which remain being integrals of motion even in an exact solution of the equations of motion. \cite{naoz2013} also showed a great agreement between the secular approximation and direct N-body integrations in most cases.

\section{The Monte Carlo Simulation}
\label{sec:GN}

In this section we introduce our method of simulating EBBHs in GNs. First, we give an overview of stellar populations residing in GNs in Section \ref{ssec:GN}. Then, we introduce the GN model of \cite{bahcall1977}, as well as we calculate the probability distributions needed for our Monte Carlo simulations (see Section \ref{ssec:Bahcall}). We also describe our Monte Carlo code for simulating EBBHs formed through gravitational capture (see Section \ref{ssec:MonteCarlo}), and introduce our approach for simulating EBBHs formed in hierarchical triple systems in Section \ref{ssec:KozaiNum}.

\subsection{Galactic nuclei and their stellar populations}
\label{ssec:GN}

The main contributors to EBBH formation are expected to be GNs, which are assumed to be spherical, relaxed stellar systems bound by gravity to a SMBH in their center. The stellar populations of GNs consist of main sequence stars (MSs), white dwarves (WDs), neutron stars (NSs), and BHs. An exhaustive investigation of relaxation around SMBHs has been carried out by various studies \citep[e.g.][]{bahcall1977,freitag2006,alexander2009,oleary2009}, who found that within a certain radius (called the \textit{radius of influence} of the SMBH), stellar objects undergo dynamical mass segregation, and form a power-law density profile ($n(r)\propto r^{-\alpha}$, where $\alpha$=$\alpha(m)$, having higher values for larger $m$ masses). The radius of influence is given by the following formula:

\begin{ceqn}
\begin{equation}
r_{\mathrm{max}} = \frac{M_{\mathrm{SMBH}}}{\sigma_*^2} ,
\label{eq:RadOfInfl}
\end{equation}
\end{ceqn}
where $M_{\mathrm{SMBH}}$ is the mass of the SMBH, and $\sigma_*$ is the velocity dispersion of the stellar system. The velocity dispersion can be obtained using the $M_{\mathrm{SMBH}}-\sigma_*$ relationship \citep{tremaine2002,norris2014}:

\begin{ceqn}
\begin{equation}
M_{\mathrm{SMBH}} \simeq 1.3 \times 10^8 \: M_{\odot} \left( \frac{\sigma_*}{200 \: \mathrm{km/s}} \right)^4 .
\label{eq:Msigma}
\end{equation}
\end{ceqn}

We used a multi-mass population of SMBHs, where the number distribution of SMBHs can be approximated as

\begin{ceqn}
\begin{equation}
\frac{\mathrm{d}N_{\mathrm{SMBH}}}{\mathrm{d}M_{\mathrm{SMBH}}} \propto \left(\frac{M_{\mathrm{SMBH}}}{M_{\bullet}}\right)^{-1.25} e^{-M_{\mathrm{SMBH}}/{M_{\bullet}}} ,
\label{eq:MSMBHdist}
\end{equation}
\end{ceqn}
where $M_{\bullet}=1.3 \times 10^8$ $M_{\odot}$, as was shown by the measurements of \cite{aller2002}. For the mass range of SMBHs, we chose a lower limit of $10^5$ $M_{\odot}$, and we set a conservative upper limit of $2 \times 10^7$ $M_{\odot}$ \citep[see][]{barth2005,greene2006,gondan2017}.

Amongst all the stellar objects in GNs, our main interests are BH populations. There is some uncertainty in the present-day mass function of stellar mass BHs in GNs, since star formation inside GNs can be quite different from the formation of field stars. Hence, there is a significant variation in the mass function suggested by models above $\sim 10$ $M_{\odot}$ \citep{alexander2009}. Similarly to previous studies \citep{alexander2009,gondan2017}, we assumed single mass MS, WD, and NS populations with component masses of \mbox{$1$ $M_{\odot}$}, $0.6$ $M_{\odot}$, and $1.4$ $M_{\odot}$, respectively. However, as it was shown by \cite{gondan2017}, the distributions of EBBH parameters are not sensitive to this choice. We chose the relative fractions of MSs, WDs, and NSs to be $1:0.1:0.01$ \citep{alexander2005}. For stellar-mass BHs, we implemented a multi-mass population, assuming a power-law mass function $m_{\mathrm{BH}}^{-\beta}$. The mass distribution of BHs is then

\begin{ceqn}
\begin{equation}
\frac{\mathrm{d}N_{\mathrm{BH}}}{\mathrm{d}m_{\mathrm{BH}}} = \frac{(1-\beta) m_{\mathrm{BH}}^{-\beta}}{m_{\mathrm{BH,max}}^{1-\beta}-m_{\mathrm{BH,min}}^{1-\beta}} ,
\label{eq:BHmass}
\end{equation}
\end{ceqn}
where $m_{\mathrm{BH,min}}$ is the lower and $m_{\mathrm{BH,max}}$ is the upper mass limit of BHs, and we chose $\beta = 2.35$ similarly to \cite{gondan2017}. Note that this $\beta$ exponent need not be related to the Salpeter initial mass function \citep{salpeter1955}. We set the minimum mass of BHs to $m_{\mathrm{BH,min}}=5$ $M_{\odot}$ based on observations of X-ray binaries \citep[see][and references therein]{belczynski2012}, and the maximum mass to $m_{\mathrm{BH,max}}=50$ $M_{\odot}$ \citep[based on][]{belczynski2016}, consistent with GW observations of binary BHs \citep{abbott2018a}. The total number of BHs in GNs is proportional to the mass of the central SMBH \citep{miralda2000}, therefore, it can be approximated as

\begin{ceqn}
\begin{equation}
N_{\mathrm{BH}} \approx 20,000 \times \frac{M_{\mathrm{SMBH}}}{M_{\mathrm{SgrA}^*}} ,
\end{equation}
\end{ceqn}
where $M_{\mathrm{SgrA}^*}=4.3 \times 10^6$ $M_{\odot}$ is the mass of the \mbox{Sgr A$^*$} SMBH \citep{gillessen2009}.

We followed the method of \cite{gondan2017} to describe EBBHs formed through gravitational capture. We used the one-body phase space distribution of \cite{bahcall1977}, generalized for multi-mass systems \citep{oleary2009} to acquire the distribution of stellar objects within relaxed GNs. We utilized these analytical distributions in our Monte Carlo simulations (see Section \ref{ssec:MonteCarlo}).

\subsection{The Bahcall \& Wolf model of galactic nuclei}
\label{ssec:Bahcall}

In our simulations we assumed relaxed GNs with spherical symmetry. Hence, we applied the one-body phase space distribution $f(\mathbf{r},\mathbf{v})$ of \cite{bahcall1977} generalized for multi-mass systems \citep{oleary2009}, which takes the following form for objects with mass $m$:

\begin{ceqn}
\begin{equation}
f_m(\mathbf{r},\mathbf{v}) = C_m E(r,v)^{p_m} ,
\end{equation}
\end{ceqn}
where $r$ and $v$ denote the magnitudes of the position vector $\mathbf{r}$ and the velocity vector $\mathbf{v}$, respectively,

\begin{ceqn}
\begin{equation}
E(r,v) = \frac{M_{\mathrm{SMBH}}}{r} - \frac{v^2}{2}
\end{equation}
\end{ceqn}
is the Keplerian binding energy per unit mass of an object in the field of the SMBH ($E(r,v)>0$ for all objects inside a relaxed GN), and $C_m$ is a normalization constant. The $p_m$ exponent is mass-dependent, with which we are able to describe mass segregation. For all stellar objects,

\begin{ceqn}
\begin{equation}
p_m = p_0 \frac{m}{m_{\mathrm{BH,max}}} ,
\end{equation}
\end{ceqn}
where $p_0=0.5$ in our simulations, based on the Fokker-Planck simulations of \cite{oleary2009}. Hence for MSs, WDs, and NSs $p_m\approx 0$.

The 3D number density of objects with mass $m$ and at radius $r$ can be obtained from the phase space distribution function:

\begin{ceqn}
\begin{equation}
n_m(r) = \int f_m(\mathbf{r},\mathbf{v}) \mathrm{d}^3 v = n_{m} (r_{\mathrm{max}}) \left( \frac{r}{r_{\mathrm{max}}} \right)^{-\alpha_m} ,
\end{equation}
\end{ceqn}
where

\begin{ceqn}
\begin{equation}
\alpha_m = \frac{3}{2} + p_m .
\end{equation}
\end{ceqn}
For MSs of $m=1$ $M_{\odot}$, the number density at the radius of influence emerges from the $M_{\mathrm{SMBH}}-\sigma_*$ relationship of Eq. (\ref{eq:Msigma}), assuming that the stellar mass within the radius of influence of the SMBH is $2M_{\mathrm{SMBH}}$ \citep[see e.g.][]{oleary2009}:

\begin{ceqn}
\begin{equation}
n_{\mathrm{MS}}(r_{\mathrm{max}}) \simeq 1.38 \times 10^{5} \mathrm{pc}^{-3} \sqrt{\frac{10^6 M_{\odot}}{M_{\mathrm{SMBH}}}} .
\end{equation}
\end{ceqn}
The number density normalization of WD, NS and BH populations is achieved as it is described in Section 2.2 of \cite{gondan2017}.

The steady-state description of a stellar population within a GN is only valid inside a certain region, where $r_{\mathrm{min}}<r<r_{\mathrm{max}}$. The maximum radius is the radius of influence of the SMBH defined by Eq. (\ref{eq:RadOfInfl}). Inside a certain region, called the ``loss cone'' \citep{Shapiro1976,Syer1999}, stellar objects get disrupted by the central SMBH and eventually merge into it. Hence $r_{\mathrm{min}}$ is set by the outer boundary of this loss cone, where the density cusp exhibits a cutoff. We calculated $r_{\mathrm{min}}$ according to the assumptions of \cite{gondan2017}, requiring that (i) the number distribution of stellar populations reaches a relaxed profile within the age of the galaxy; and (ii) the relaxation time is shorter than the timescale of inspiral into the SMBH \citep[see Appendix A of][]{gondan2017}.

The velocity distribution of stellar objects can also be obtained from the phase space distribution function:

\begin{ceqn}
\begin{equation}
\varphi(\bar{v},m) = C_{\bar{v}} (1-\bar{v}^2)^{p_m} ,
\label{eq:veldist}
\end{equation}
\end{ceqn}
where $\bar{v} = v/v_{\mathrm{max}}(r)$ is the dimensionless velocity, with $v_{\mathrm{max}}$ being the escape velocity at radius $r$, and $C_{\bar{v}}$ is a normalization factor. In our Monte Carlo simulations, one of the steps was to generate randomized values of relative velocities, hence constructing the distribution of relative velocities was essential. Using the velocity distributions $f_1(\mathbf{v}_1)$ and $f_2(\mathbf{v}_2)$ of objects $1$ and $2$, the distribution of the magnitude of relative velocities $w$ is

\begin{equation*}
F_{1,2}(w) = \int \mathrm{d}^3v_1 f_1(\mathbf{v}_1)
\end{equation*}
\vspace{-0.5cm}
\begin{ceqn}
\begin{equation}
\times \int \mathrm{d}^3v_2 f_2(\mathbf{v}_2) \delta(w-|\mathbf{v}_1-\mathbf{v}_2|) ,
\end{equation}
\end{ceqn}
where $\delta(\cdot)$ is the Dirac-delta function. Using Eq. (\ref{eq:veldist}), the distribution of the magnitude of relative velocities in the Bahcall \& Wolf model can be expressed with a single integral \citep[see Appendix B of][]{gondan2017}:

\begin{equation*}
F_{1,2}(\bar{w}) = \frac{4 \pi^2 \bar{w}}{p_2 + 1}
\end{equation*}
\vspace{-0.5cm}
\begin{ceqn}
\begin{equation}
\times \int\limits_{\bar{w}-1}^1 \bar{v}_1 (1-\bar{v_1}^2)^{p_1} \left[ 1-(\bar{v}_1 - \bar{w})^2 \right]^{p_2+1} \mathrm{d}\bar{v}_1 ,
\end{equation}
\end{ceqn}
where $\bar{w} = w/v_{\mathrm{max}}(r)$ ($0 \leq \bar{w} \leq 2$), and $p_1$ and $p_2$ are the $p_m$ exponents corresponding to $m_1$ and $m_2$, respectively. The remaining integral can be evaluated analytically for any integer or half-integer $p_1$ and $p_2$, although in a general case it can only be calculated numerically.

\subsection{EBBH formation rates and the Monte Carlo method}
\label{ssec:MonteCarlo}

We simulated the formations of EBBHs through gravitational capture using a Monte Carlo code. In our simulation, we used analytical expressions for EBBH formation rates. The infinitesimal formation rate (measured in s$^{-1}$ units) for a fixed $M_{\mathrm{SMBH}}$ is

\begin{ceqn}
\begin{equation}
\mathrm{d}^{9}\Gamma_{1,2} = \sigma w f_1(\mathbf{r},\mathbf{v}_1) f_2(\mathbf{r},\mathbf{v}_2) \, \mathrm{d}^3r \, \mathrm{d}^3v_1 \, \mathrm{d}^3v_2 ,
\end{equation}
\end{ceqn}
where $1$ and $2$ index the two BHs in the EBBH, and $\sigma = \sigma(w) = \pi [b^2_{\mathrm{max}}(w)-b^2_{\mathrm{min}}(w)]$ is the cross section for BHs forming an EBBH through gravitational capture (see Eq. (\ref{eq:bminmax}) for $b_{\mathrm{min}}$ and $b_{\mathrm{max}}$). We used the results of \cite{gondan2017}, who carried out the analytical calculations to determine the distribution of EBBH formation rate as a function of different parameters. Compared to the simulations of \cite{gondan2017}, we expanded our parameter space by picking the $M_{\mathrm{SMBH}}$ values randomly too. The formation rate as a function of $M_{\mathrm{SMBH}}$ can be calculated as

\begin{figure}
	\includegraphics[width=\columnwidth]{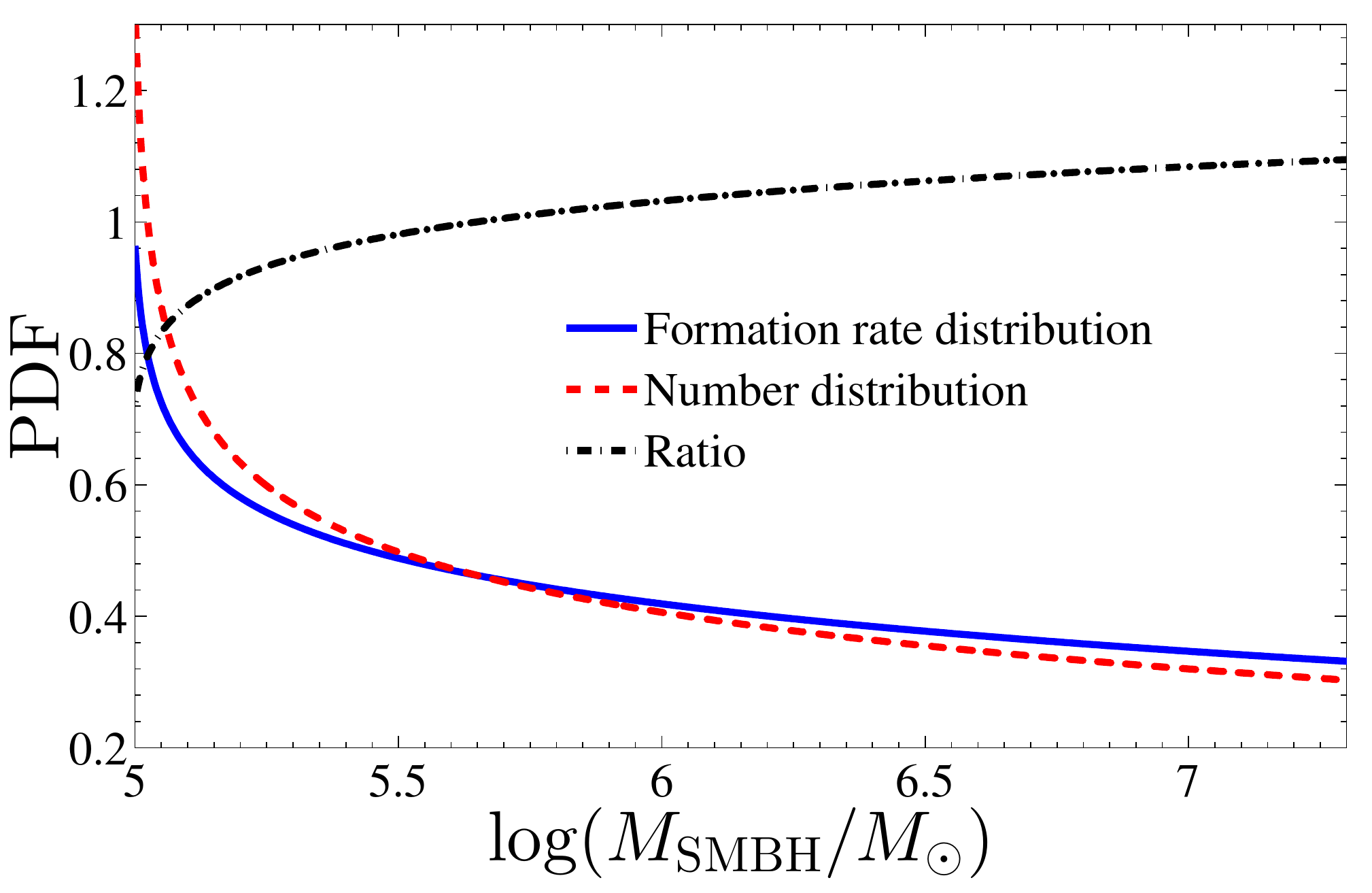}
    \caption{The probability density function (PDF), as a function of SMBH mass $M_{\mathrm{SMBH}}$, of an EBBH forming through gravitational capture in a GN with $M_{\mathrm{SMBH}}$ in its centre (solid blue line). We obtained this PDF using analytical calculations described in Section \ref{ssec:MonteCarlo}. We also show the PDF of $M_{\mathrm{SMBH}}$ masses in GNs given by Eq. (\ref{eq:MSMBHdist}) (dashed red line), as well as the ratio of the two (dash-dotted black line). Note, that the ratio is proportional to the formation rate of EBBHs in a single GN. The ratio increases only slightly for higher SMBH masses \citep[see Appendix B of][for further discussion]{gondan2017}.}
    \label{fig:MsmbhEv}
\end{figure}

\begin{ceqn}
\begin{equation}
\left\langle \frac{\partial \Gamma}{\partial M_{\mathrm{SMBH}}} \right \rangle =  \frac{\mathrm{d}N_{\mathrm{SMBH}}}{\mathrm{d}M_{\mathrm{SMBH}}} \left\langle \Gamma \right \rangle_{M_{\mathrm{SMBH}}} ,
\end{equation}
\end{ceqn}
where $\left\langle \Gamma \right \rangle_{M_{\mathrm{SMBH}}}$ is the formation rate of EBBHs in a single GN with $M_{\mathrm{SMBH}}$ in its centre. We show the distribution of the EBBH formation rate (red dashed line) and the SMBH number distribution (blue solid line) as functions of $M_{\mathrm{SMBH}}$ in Fig. \ref{fig:MsmbhEv}. Note that the formation rate distribution is mainly determined by the mass distribution of SMBHs and is only weakly affected by the $M_{\mathrm{SMBH}}$ dependency of the formation rate of EBBHs in a single GN (see the black dash-dotted line in \mbox{Fig. \ref{fig:MsmbhEv}}). A thorough investigation on the $M_{\mathrm{SMBH}}$ dependence of the formation rate of EBBHs in a single GN gives $\left\langle \Gamma \right \rangle_{M_{\mathrm{SMBH}}} \propto M_{\mathrm{SMBH}}^{\gamma}$ with $\gamma \approx 0.1 - 0.25$ for single-mass BH populations with different BH masses \citep[see Appendix B of][]{gondan2017}.

In our Monte Carlo simulation, we applied the following steps. First, we used the EBBH formation rate distribution shown in  Fig. \ref{fig:MsmbhEv} to generate a number of random $M_{\mathrm{SMBH}}$ values, thus simulating GNs with different SMBH masses. Then for each $M_{\mathrm{SMBH}}$, we drew an $m_1$-$m_2$ pair from their corresponding two-dimensional distribution, where both masses range from $5$ $M_{\odot}$ to $50$ $M_{\odot}$ \citep[see Appendix B.3 in][for details]{gondan2017}. For each $m_1$-$m_2$ pairs, we randomly drew $r$ radii values from the range [$r^{1,2}_{\mathrm{min}},r_{\mathrm{max}}$] \cite[see Eq. (\ref{eq:RadOfInfl}) and Appendix A of][for details]{gondan2017}. Lastly, for each $r$ we randomly drew pairs of $w$ and $b$ values from ranges [$0,2 v_{\mathrm{max}}$] and [$b_{\mathrm{min}},b_{\mathrm{max}}$], respectively.

The analytical estimates we used in our simulations can be violated by EBBHs that interact with a third object between their formation and merger. \cite{gondan2017} showed that this affects only a small fraction ($<1$ percent) of binaries over the considered ranges of parameters. EBBHs can form with gravitational capture if the impact parameter of the encounter between the BHs is smaller than $b_{\mathrm{max}}$ (so that enough energy is emitted during the encounter for the binary to form), and larger than $b_{\mathrm{min}}$  (to avoid the direct collision of the BHs). These parameters scale as $b_{\mathrm{min}} \propto w^{-1}$ and $b_{\mathrm{max}} \propto w^{-9/7}$ with the relative velocity of the BHs. As a consequence, in the inner regions of GNs, near $r_{\mathrm{min}}$, the condition $b_{\mathrm{max}} > b_{\mathrm{min}}$ can be violated. We discarded all such $(w,b)$ pairs, however, as it was shown by \cite{gondan2017}, this corresponds only to a small fraction ($<1$ percent) of binaries over the ranges of parameters we considered in our simulations.

\subsection{Numerical approach for hierarchical triples}
\label{ssec:KozaiNum}

The formation of hierarchical triple systems of BHs is hardly manageable with analytical calculations, since they can form through various different processes, most of which include multiple interactions between binary systems that can easily become untraceable from an analytical point of view. Therefore, we applied a Monte Carlo approach different from the one employed for gravitational captures. We adopted a set of distributions to set the initial parameters of triple systems. We then used these initial conditions as inputs for our differential equation integrator, and solved the octupole-order equations of motion of \cite{naoz2013}, in which we included gravitational radiation. For solving the differential equations we utilized a fourth-order Runge-Kutta differential equation integrator with adaptive stepsize control to increase the speed of the integration while not losing precision.

We picked the masses of BHs making up the inner binary independently from the distribution defined by Eq. (\ref{eq:BHmass}) using an exponent of $\beta=2.35$. The third object in the triple system was always chosen as the central SMBH. The mass of the SMBH was picked from the distribution defined by Eq. (\ref{eq:MSMBHdist}). The lower and upper limits for masses were chosen to be the same as for gravitational captures. The initial semi-major axis of the inner binary, $a_1$, was drawn uniformly in the logarithmic domain (which corresponds to d$N/$d$a\propto 1/a$). The minimum and maximum values of $a_1$ were chosen as $0.1$ AU and $50$ AU, respectively, similarly to the setup of \cite{hoang2018}. The initial semi-major axis of the outer binary was drawn from a uniform distribution (which corresponds to a 3D number density distribution of $n(r)\propto r^{-2}$) between \mbox{$100$ AU} and $0.1$ pc, based on the simulational setup of \cite{hoang2018}. Initial eccentricities of the inner and outer binaries were drawn between $0$ and $1$ from uniform \citep{raghavan2010}, and thermal distributions \citep[d$N/$d$e \propto e$, see][]{jeans1919}, respectively. $g_1$ and $g_2$ were picked randomly from uniform distributions, while the initial inclination $I$ was drawn from a distribution d$N/$d$I\propto \cos(I)$ over the range of $[65^{\circ},125^{\circ}]$. The range of initial inclination was based on the results of \cite{silsbee2017}, who found that mergers typically occur with $-0.5<\cos(I)<0.5$.

Since we use secular approximations when describing triple systems, we require our systems to satisfy dynamical stability in order to justify the hierarchical treatment. Hence, we use two stability criteria \citep[see e.g.][]{stephan2016,hoang2018}. First we require that the relative strengths of the octupole and quadrupole terms fulfil the following criterion \citep{naoz2016}:

\begin{ceqn}
\begin{equation}
\epsilon = \frac{a_1}{a_2}\frac{e_2}{1-e_2^2} < 0.1 .
\end{equation}
\end{ceqn}
The second condition requires that the inner binary does not cross the SMBH's Roche limit \citep[e.g.][]{naozsilk2014}:

\begin{ceqn}
\begin{equation}
\frac{a_2}{a_1} > \left(\frac{3 M_{\mathrm{SMBH}}}{M}\right)^{1/3} \  \frac{1+e_1}{1-e_2} ,
\end{equation}
\end{ceqn}
where $M=m_1+m_2$ is the total mass of the inner binary. We discarded all triple systems that failed to satisfy these conditions.

The highly eccentric excitations can be inhibited by the general relativistic precession of the inner binary, as was shown by \cite{naozkocsis2013}. This happens when the timescale of Kozai-Lidov oscillations at the quadrupole level of approximation \citep[$t_\mathrm{quad}$, see e.g.][]{antognini2015} is much longer than the timescale of general relativistic precession \citep[$t_\mathrm{GR,inner}$, see e.g.][]{naozkocsis2013}. We consider this phenomenon in our simulations by discarding all triple systems for which $t_\mathrm{GR,inner} < t_\mathrm{quad}$, as the inner binaries of these systems may not experience highly eccentric excitations.

We terminated our simulation if at any time $a_1$ became smaller than $10^{-4}$ AU, which we assumed to guarantee a merger within a short period of time. Close encounters with other stellar objects inside the GN can unbind the inner binary during its evolution. This evaporation timescale is

\begin{ceqn}
\begin{equation}
t_{\rm ev} = \frac{\sqrt{3}\sigma}{32\sqrt{\pi} G\rho a_1\ln \Lambda}\frac{M}{m_\mathrm{ave}} ,
\end{equation}
\end{ceqn}
where in the inner parts ($\leq 0.1$ pc)

\begin{ceqn}
\begin{equation}
\rho= \frac{3}{4\pi}\frac{M_{\rm SMBH}}{a_2^3}\left( \frac{G\sqrt{M_{\rm SMBH} M_0}}{\sigma_0^2 a_2}\right)^{-3/2}
\end{equation} 
\end{ceqn}
is the density of stars, $m_\mathrm{ave} = 1$ $M_\odot$ is the average mass of background stars, $\ln \Lambda = 15$ is the Coulomb logarithm, $\sigma = \sqrt{0.1 \: \mathrm{pc} / a_2}$ $280$ km/s is the velocity dispersion, $G$ is the universal gravitational constant, while $\sigma_0=200$ km/s and $M_0 = 3 \times 10^8$ $M_\odot$ are constants \citep[see][]{hoang2018}. We simply terminated our simulation and discarded the triple system if the inner binary has not produced a merger after $t_{\mathrm{ev}}$. The inner binary can merge without experiencing high-eccentricity excitations. However, we are only interested in those mergers that take place due to Kozai-excitations. Therefore we discarded those triple systems, in which the merger did not happen due to high-eccentricity excitations. In addition to these conditions, we kept only those EBBHs, for which effects due to GW emission dominated over secular effects at \mbox{$f_{\mathrm{GW}} = 10$ Hz}, and this condition held throughout their evolution from that point onward (see Section \ref{ssec:RecPar} for details).

\section{Results}
\label{sec:KozaiGN}

In this section we present our results. First, we discuss how we acquired the distributions of EBBH eccentricities at $f_{\mathrm{GW}}=10$ Hz for all formed EBBHs by evolving EBBH orbits in our simulations described in Section \ref{ssec:MonteCarlo} and \ref{ssec:KozaiNum}. We then construct the distributions of eccentricities for future EBBH observations with design aLIGO detectors (see Section \ref{ssec:RecPar}). In Section \ref{ssec:KozaiRes} and \ref{ssec:KozaiResErr} we also show that if future EBBH detection rates with aLIGO will be dominated by EBBHs formed in GNs, then it is feasible to constrain the branching ratios between the two main channels. We performed a statistical test to determine the minimum number of EBBH observations needed to constrain the branching ratio of gravitational capture to an arbitrarily chosen $\Delta \mathcal{F}\mathrm{_{GC}} = 0.2$ wide one-sigma confidence interval (note that due to the underlying symmetry, our results would be the same if the branching ratio for the secular Kozai-Lidov mechanism was constrained instead). We present our results without accounting for errors of the reconstruction of EBBH orbital parameters in Section \ref{ssec:KozaiRes}. We then discuss the implications of non-zero parameter reconstruction errors in \mbox{Section \ref{ssec:KozaiResErr}}.

\subsection{Eccentricity distributions at $f_{\mathrm{GW}}=10$ Hz}
\label{ssec:RecPar}

We produced the distributions of eccentricities of EBBHs at the time their peak GW frequency, $f_{\mathrm{GW}}$, reaches \mbox{$10$ Hz}, from now on denoted as $e_{\mathrm{LIGO}}$. For EBBHs formed through gravitational capture, we produced the $e_{\mathrm{LIGO}}$ distribution for all formed EBBHs by solving Eqs. (\ref{eq:rhoe}) and (\ref{eq:fgw}) for \mbox{$f_{\mathrm{GW}}=10$ Hz}, using initial orbital parameters obtained from our simulation described in Section \ref{ssec:MonteCarlo}. We also acquired the $e_{\mathrm{LIGO}}$ distribution for EBBHs formed through the Kozai-Lidov mechanism from our simulations described in Section \ref{ssec:KozaiNum}. Up until the last few Kozai-cycles before the merger, the evolution of the inner binary is dominated by perturbing effects of the outer object, after which the domination is taken over by the energy and angular momentum loss due to gravitational radiation (see \mbox{Fig. \ref{fig:KozaiQPOP}} as an example). According to our results, the proportion of EBBHs formed through the Kozai-Lidov mechanism, for which secular effects still cannot be ignored at the time their GW signals enter the aLIGO band is $<1$ percent. For such EBBHs, both finding the GW signal and reconstructing EBBH parameters pose a complex problem. Thus, for simplicity, we leave out these EBBHs from our analysis.

The distribution of physical parameters is different for detected EBBHs than for formed EBBHs, hence we needed to produce the $e_{\mathrm{LIGO}}$ distribution for all detected EBBHs in order to carry out our analysis. This is not equivalent with the $e_{\mathrm{LIGO}}$ distribution of all formed EBBHs, which must be weighted by the volume from which EBBHs can be observed, and the maximum distance to which EBBHs are observable (called the horizon distance, which we denote by $D_{\mathrm{H}}$) differs for EBBHs with different component masses and eccentricities. \cite{oleary2009} showed that BH binaries with higher initial eccentricities are detectable only to slightly larger distances than circular binaries for low-mass (\mbox{$M<60$ $M_{\odot}$}) binaries, and the difference in the horizon distance between circular and eccentric binaries becomes substantial only at higher masses (\mbox{$M\approx100$ $M_{\odot}$}). For EBBHs formed through the secular Kozai-Lidov mechanism the proportion of EBBHs with $M>60$ $M_{\odot}$ is only $\sim1$ percent. Therefore for simplicity, for EBBHs formed through this channel, we defined $D_{\mathrm{H}}$ the same way as it is defined for circular binaries \citep[see e.g.][]{abadie2010}:

\begin{ceqn}
\begin{equation}
D_{\mathrm{H}} = \frac{1}{8} \left( \frac{5\pi}{6} \right)^{1/2} \mathcal{M}^{5/6} \pi^{-7/6} \sqrt{\int_{f_{\mathrm{low}}}^{f_{\mathrm{high}}} \frac{f^{-7/3}}{S_n(f)} \mathrm{d}f} ,
\label{eq:HDist}
\end{equation}
\end{ceqn}
where

\begin{figure}
\centering
	\includegraphics[width=0.7\columnwidth]{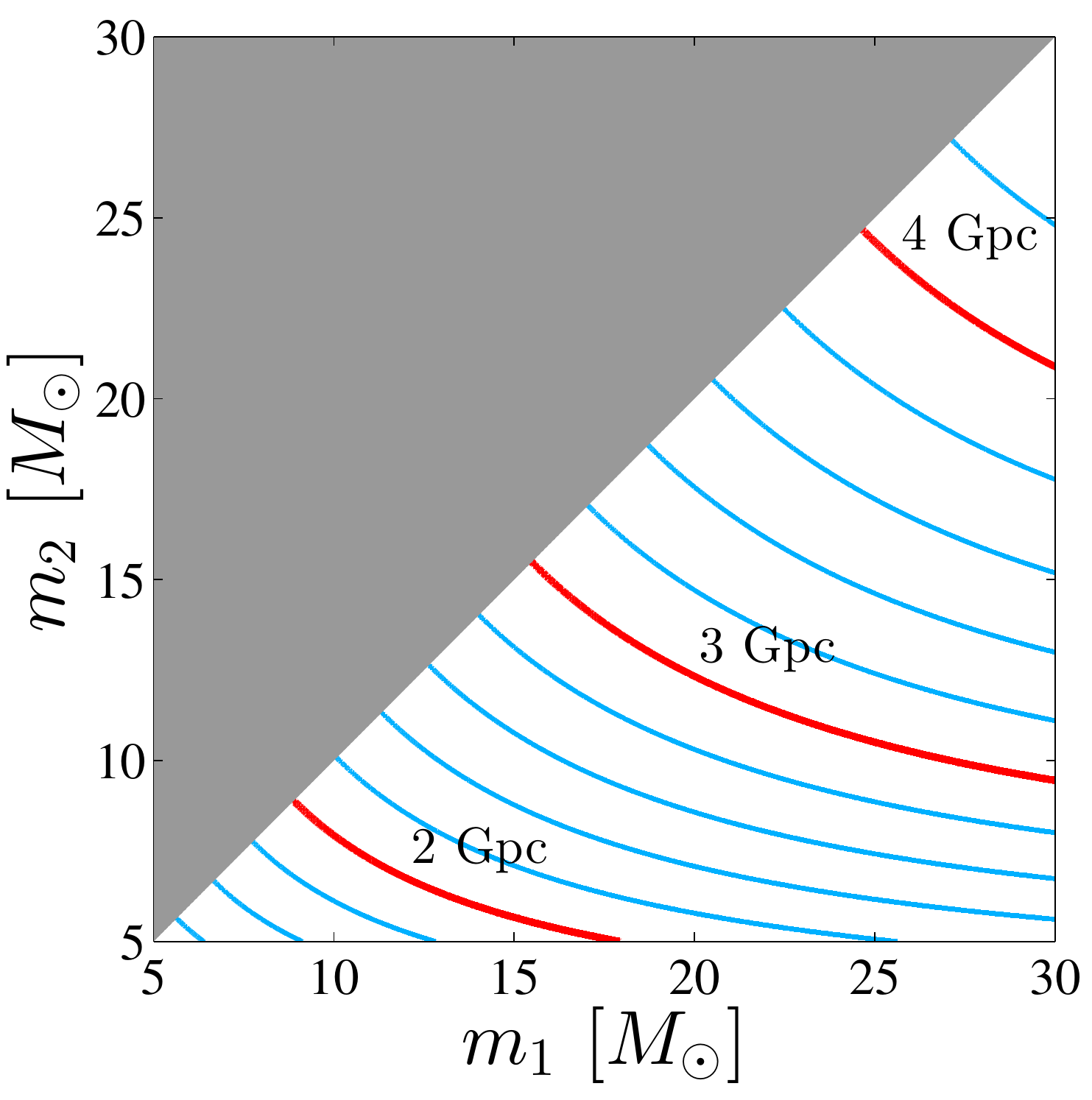}
	\caption{Horizon distances $D_{\mathrm{H}}$ of a single aLIGO detector with design sensitivity for circular binary BHs, as a function of component masses $m_1$ and $m_2$ (defined as $m_1 \geq m_2$). Different contours show $D_{\mathrm{H}}$ increasing in steps of $1$ Gpc (thick red lines) and $0.2$ Gpc (thin blue lines). Horizon distances for EBBHs with total masses of \mbox{$M<60$ $M_\odot$} differ only slightly from that of circular binaries (see Section \ref{ssec:RecPar} for details).}
    \label{fig:HDist}
\end{figure}

\begin{ceqn}
\begin{equation}
\mathcal{M} = \frac{(m_1 m_2)^{3/5}}{(m_1 + m_2)^{1/5}}
\end{equation}
\end{ceqn}
is the chirp mass, $f_{\mathrm{low}}$ is set by the detector noise cutoff (which in our case is $f_{\mathrm{low}}=10$ Hz), \mbox{$f_{\mathrm{high}} = (6 \sqrt{6} \pi M)^{-1}$} is the frequency at the innermost stable circular orbit, and $S_n(f)$ is the power spectral density of the detector noise, for which we used the design aLIGO projection \citep[see][]{aasi2015a}. For EBBHs formed through gravitational capture, we used the horizon distance calculated for eccentric binaries in Eq. (58) of \cite{oleary2009}, since in this channel the proportion of binaries with total masses of \mbox{$M>60$ $M_\odot$} is $\sim10$ percent. The number of detectable EBBHs with certain parameter values should be proportional to the volume enclosed by the corresponding horizon distance. Hence, to convert the $e_{\mathrm{LIGO}}$ distribution of all formed EBBHs to the distribution of all detected EBBHs, we weighted each EBBH parameter set by $V_{\mathrm{H}}=4\pi D_{\mathrm{H}}^3 /3$. The horizon distance $D_{\mathrm{H}}$ as a function of BH masses for low-mass binaries ($m_2 \leq m_1 \leq 30$ $M_\odot$, defined by Eq. (\ref{eq:HDist})) is shown in \mbox{Fig. \ref{fig:HDist}}. The binaries with the smallest masses are detectable to \mbox{$\sim 1.3$ Gpc}, while binaries with $M=60$ $M_\odot$ have a corresponding horizon distance of $\sim 4.4$ Gpc.

\begin{figure*}
	\includegraphics[width=0.49\textwidth]{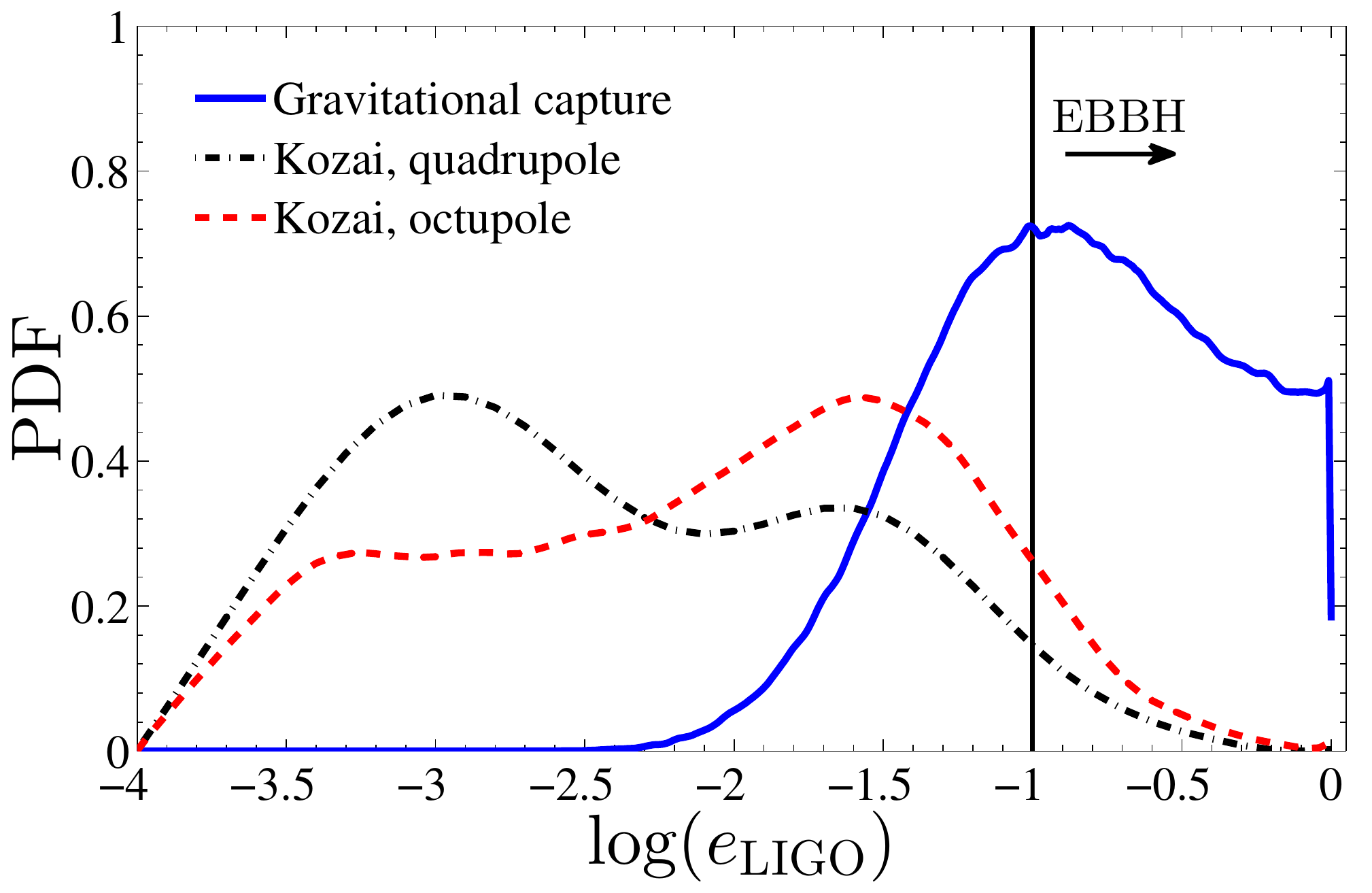}
	\includegraphics[width=0.49\textwidth]{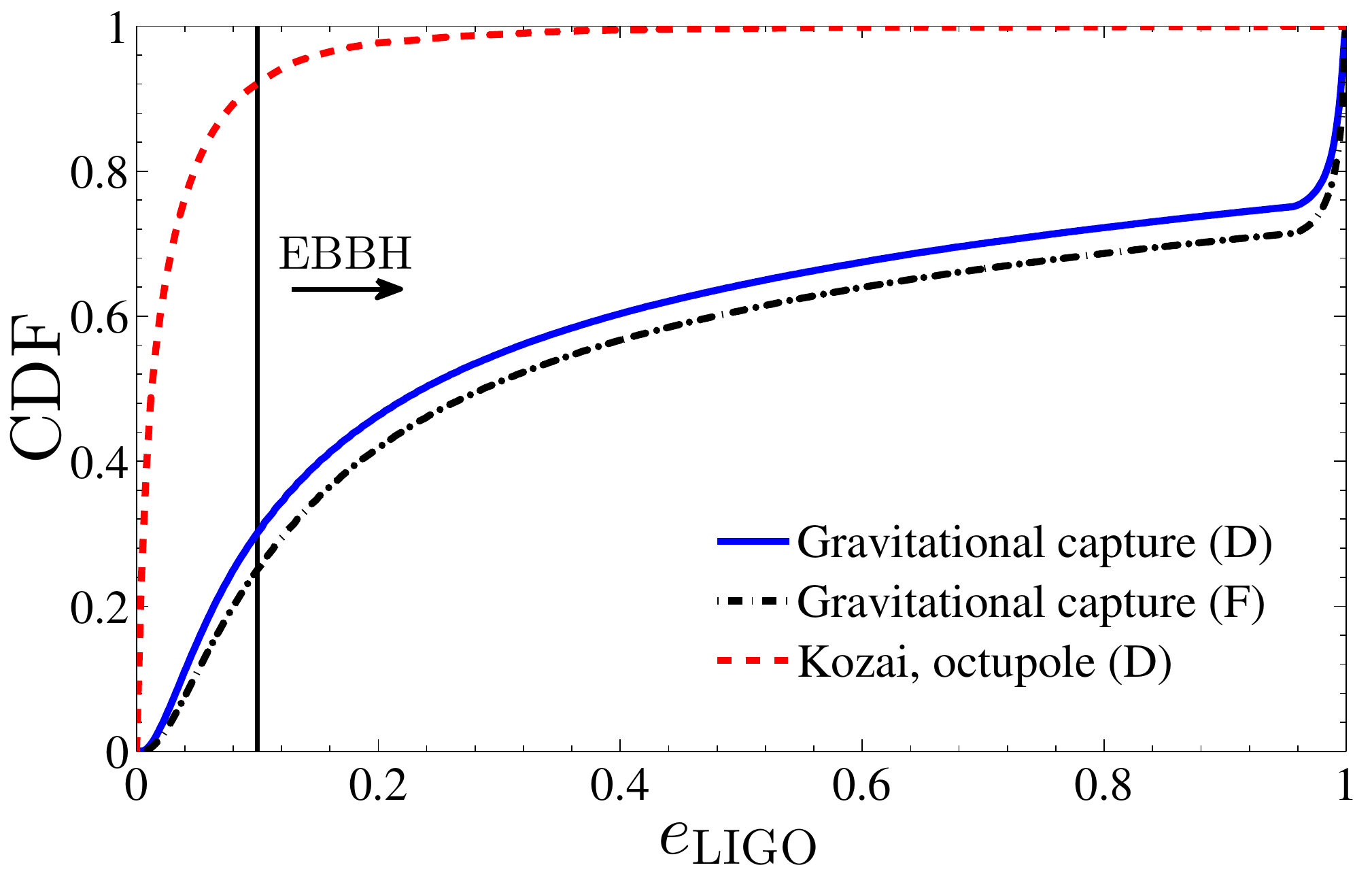}
    \caption{Simulated distributions of $e_{\mathrm{LIGO}}$ for detected (D) and for formed (F) binary BHs (including EBBHs), corresponding to gravitational captures and the secular Kozai-Lidov mechanism as formation channels in GNs. The left panel shows the smoothed PDFs of $e_{\mathrm{LIGO}}$ for detected EBBHs formed through gravitational capture with $f_{\mathrm{GW,0}}<10$ Hz, as well as for EBBHs formed through the secular Kozai-Lidov mechanism, for which we both show the PDFs resulting from the quadrupole-, and octupole-order equations of motion. As a consequence of using the octupole-order approximation, which results in EBBHs experiencing extreme Kozai-excitations, the corresponding distribution generally shifts to higher eccentricities. The right panel shows the cumulative distribution functions (CDFs) for detected EBBHs, as well as for formed EBBHs in the case of EBBHs formed through gravitational capture. The CDFs of formed and detected EBBHs for the secular Kozai-Lidov mechanism overlap within 5 percent.}
    \label{fig:eLigo}
\end{figure*}

The $e_{\mathrm{LIGO}}$ distributions of detected EBBHs for the two different formation channels are shown in Fig. \ref{fig:eLigo}. For EBBHs formed through the secular Kozai-Lidov mechanism, we show the $e_{\mathrm{LIGO}}$ probability distribution functions resulting from both the quadrupole- and octupole-order equations of motion. Using the octupole-order equations generally produces EBBHs with higher eccentricities. This is understandable, since we expect EBBHs to begin their inspiral from higher initial eccentricities due to extreme Kozai-excitations caused by the octupole term. Note that the $e_{\mathrm{LIGO}}$ distributions for detected EBBHs only slightly differ from the corresponding distributions for all formed EBBHs (see the right panel of Fig. \ref{fig:eLigo}). We also find, supplementing the results of \cite{hoang2018}, that about 10 percent of binary BHs formed through the Kozai-Lidov mechanism in GNs have eccentricities $e_{\mathrm{LIGO}}>0.1$ \citep[note that the range and distribution of BH masses in our simulations were different from that of][and we did not use explicitly the 1PN terms describing orbital precession, these, however, should not change the percentage value significantly]{hoang2018}. 

For EBBHs formed through gravitational capture two main groups can be distinguished in the cumulative distribution function (see the right panel of \mbox{Fig. \ref{fig:eLigo}}). More than half of the EBBHs form outside the sensitive band of aLIGO detectors, with the initial frequency of the GW signal being \mbox{$f_{\mathrm{GW,0}}<10$ Hz}. The eccentricities of these binaries at $f_{\mathrm{GW}}=10$ Hz fall between $10^{-2}<e_{\mathrm{LIGO}}<1$. The other group consists of EBBHs that form inside the sensitive band of aLIGO detectors, with $f_{\mathrm{GW,0}} \geq 10$ Hz. These binaries form a sharp peak in the probability distribution function at eccentricities \mbox{$e_{\mathrm{LIGO}}>0.95$}. For binary BHs formed through gravitational capture in GNs, we find that about 75 percent have eccentricities $e_{\mathrm{LIGO}}>0.1$. Table \ref{tab:gondan} gives further information about the fraction of EBBHs under various conditions \citep[similarly to Table 1 in][]{gondan2017}.

EBBHs tend to have higher eccentricities at lower signal frequencies. Hence we may also be able to constrain these formation channels using future GW observations with detectors operating at lower frequencies (i.e. with LISA or DECIGO). However, as it was recently shown by several studies, binary black holes that enter the aLIGO band with $e>5\times10^{-3}$ may completely elude the LISA band in the epoch their evolution is dominated by GW emission \citep[see e.g.][and references therein]{antonini2017,samsing2018,dorazio2018}. EBBHs formed in hierarchical triples may still enter the LISA band multiple times during the time secular evolution dominates over gravitational radiation. The parameter reconstruction of these systems, however, pose a more complex problem than in the case of an EBBH in its merger phase. The proportion of EBBHs formed through the secular Kozai-Lidov mechanism, for which gravitational radiation does not dominate over secular effects of the outer object in the DECIGO band, may also be considerably higher.

\begin{table}
 \caption{Fraction of EBBHs having $e_\mathrm{LIGO}>0.1$ at \mbox{$f_\mathrm{GW}=10$ Hz} ($F_{e_\mathrm{LIGO}>0.1}$), of EBBHs starting their GW-dominated evolution within the aLIGO band ($F_{\mathrm{LIGO}}$), and of EBBHs having eccentricities larger than $0.1$ at their last stable orbit ($F_{e_\mathrm{LSO}>0.1}$). The table rows correspond to the gravitational capture (GC) and the secular Kozai-Lidov (KL) formation channels, respectively.}
 \label{tab:gondan}
\begin{center}
 \begin{tabular}{lccc}
  \hline
  \hline
  Formation & $F_{e_\mathrm{LIGO}>0.1}$ & $F_{\mathrm{LIGO}}$ & $F_{e_\mathrm{LSO}>0.1}$\\
  \hline
  GC & $75\%$ & $28\%$ & $11\%$ \\[2pt]
  KL & $11\%$ & $<1\%$ & $<1\%$ \\
  \hline
 \end{tabular}
\end{center}
\end{table}

The $e_{\mathrm{LIGO}}$ distributions of detected EBBHs determined in this section are idealized forms of the distributions of observed $e_{\mathrm{LIGO}}$ values, since they do not take into account the potential errors of the parameter estimation. We considered a set of plausible errors consistent with the results of previous studies \citep[see][]{gondan2018,lower2018,gondankocsis2018} to account for the magnitudes of errors on the parameter estimations in our statistical tests (see Section \ref{ssec:KozaiResErr}).

\subsection{Constraining EBBH formation channels with zero parameter estimation errors}
\label{ssec:KozaiRes}

While this may not be true (see Section \ref{sec:Introduction} for further explanations), in this section and in Section \ref{ssec:KozaiResErr} we assume that future EBBH detection rates with aLIGO will be dominated by EBBHs formed in GNs. We perform a statistical test to demonstrate the feasibility of constraining the branching ratios between gravitational capture and the secular Kozai-Lidov mechanism in GN host environments, when only these formation channels are taken into account. To carry out our analysis, we needed to produce the distributions of physically measurable EBBH parameters. In principle, reconstructible parameters of non-spinning EBBHs include sky location, BH masses, spins, and orbital parameters ($e$, $\rho_{\mathrm{p}}$) at a given point of EBBH evolution. In realistic scenarios, however, the reconstruction of these parameters is liable to a number of limitations. First of all, the reconstruction of EBBH parameters may involve degeneracies. For example, for EBBHs formed through gravitational capture, initial orbital parameters may only be reconstructed if the peak GW frequency of the initial orbit is large enough to be in the sensitive frequency band of GW detectors \citep[see][]{gondan2018}. Parameter estimation errors are also inherently involved in the reconstruction due to the finite sensitivity of GW detectors \citep[see e.g.][]{becsy2017}. These errors may also be affected by theoretical uncertainties of the waveform model \citep{cutler2007}. \cite{kocsislevin2012} showed that post-Newtonian calculations of different order lead to substantially different orbital evolutions at small separations ($\rho_p < 20$), although they do not deviate considerably from the leading order results for higher separations. On the other hand, as already mentioned in Section \ref{ssec:RecPar}, the reconstruction of parameters of EBBHs formed through the secular Kozai-Lidov mechanism may only be accurate if the inner binary has achieved a low enough separation to have its evolution governed by gravitational radiation instead of secular effects.

Considering these factors, we chose the parameter most suitable for our analysis to be $e_{\mathrm{LIGO}}$, since at \mbox{$f_\mathrm{GW}=10$ Hz} the two BHs have a high enough separation that the post-Newtonian calculations are still accurate, while they have already shrunk to an orbit, where secular effects caused by a third object can be neglected. We chose the distribution of an EBBH orbital parameter over distributions of BH masses, since distributions of orbital parameters only weakly depend on the choice of the distribution of BH masses \citep[see][]{gondan2017}. The results of \cite{kocsislevin2012} suggest that the choice of \mbox{$f_{\mathrm{GW}}=10$ Hz} corresponds to a point of EBBH evolution where the leading-order approximations are still highly accurate. On the other hand, we kept only those EBBHs in hierarchical triples, for which effects due to GW emission dominated over secular effects at $f_{\mathrm{GW}} = 10$ Hz, and this condition held throughout their evolution from that point onward.

\begin{figure}
	\includegraphics[width=\columnwidth]{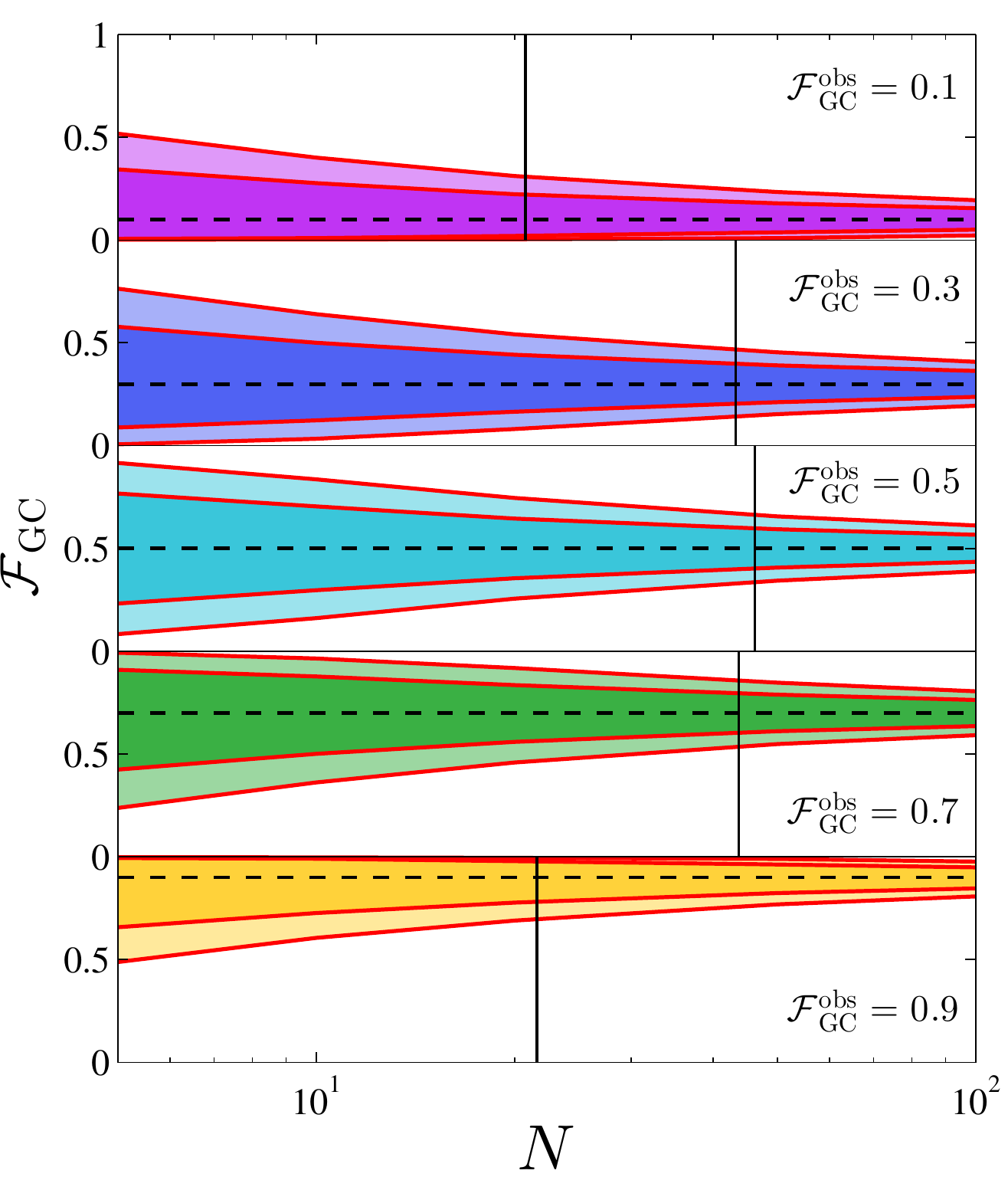}
    \caption{The 68 percent (dark shading) and 90 percent (light shading) confidence intervals of $\mathcal{F}\mathrm{_{GC}}$ distributions as functions of the number of future EBBH observations with design aLIGO ($N$), and for different values of $\mathcal{F}\mathrm{_{GC}^{obs}}$. Note that here we neglected all errors on EBBH parameter estimations. The distributions of $\mathcal{F}\mathrm{_{GC}}$ converge to $\mathcal{F}\mathrm{_{GC}^{obs}}$ (marked by dashed horizontal lines) with increasing $N$ in each case. The plots for different values of $\mathcal{F}\mathrm{_{GC}^{obs}}$ are symmetric to $\mathcal{F}\mathrm{_{GC}^{obs}}=0.5$ (which means that the plots would practically be the same if instead of gravitational capture, the secular Kozai-Lidov mechanism was considered). The widths of the 68 percent confidence intervals reach $\Delta \mathcal{F}\mathrm{_{GC}} = 0.2$ at $N=21$ for $\mathcal{F}\mathrm{_{GC}^{obs}}=0.1$ (and for $\mathcal{F}\mathrm{_{GC}^{obs}}=0.9$), at $N=43$ for $\mathcal{F}\mathrm{_{GC}^{obs}}=0.3$ (and for $\mathcal{F}\mathrm{_{GC}^{obs}}=0.7$), and at $N=46$ for $\mathcal{F}\mathrm{_{GC}^{obs}}=0.5$. These $N$ values are marked with solid vertical lines.}
    \label{fig:KozaiGNADTest}
\end{figure}

We carried out our statistical test in the following manner. We constructed a mixed distribution of $e_{\mathrm{LIGO}}$ values by weighting the $e_{\mathrm{LIGO}}$ PDF for gravitational captures (see \mbox{Fig. \ref{fig:eLigo}}) by a factor of $\mathcal{F}\mathrm{_{GC}}$, and the $e_{\mathrm{LIGO}}$ PDF for the secular Kozai-Lidov mechanism by a factor of $1-\mathcal{F}\mathrm{_{GC}}$, and then summing them up together. We then picked $N$ number of $e_{\mathrm{LIGO}}$ values randomly from this mixed distribution, and maximized the likelihood of this $N$-element sample over all possible values of $\mathcal{F}\mathrm{_{GC}} \in [0,1]$ using the Anderson-Darling test \citep{anderson1952}. We denote the resulting $\mathcal{F}\mathrm{_{GC}}$ value as $\mathcal{F}\mathrm{_{GC}^{obs}}$. We repeated this process $10^5$ times for a given $N$, to map the PDF of $\mathcal{F}\mathrm{_{GC}^{obs}}$ for a fixed value of $\mathcal{F}\mathrm{_{GC}}$. For each $N$ value, we performed this test for different values of $\mathcal{F}\mathrm{_{GC}}$ ranging from $\mathcal{F}\mathrm{_{GC}}=0$ to $\mathcal{F}\mathrm{_{GC}}=1$, obtaining the corresponding PDFs of $\mathcal{F}\mathrm{_{GC}^{obs}}$ values. We then determined the PDF of $\mathcal{F}\mathrm{_{GC}}$ for a fixed value of $\mathcal{F}\mathrm{_{GC}^{obs}}$ using conditional probability, and determined the 68 percent and 90 percent confidence intervals of these distributions for different values of $\mathcal{F}\mathrm{_{GC}^{obs}}$ ranging from $\mathcal{F}\mathrm{_{GC}^{obs}}=0.1$ to $\mathcal{F}\mathrm{_{GC}^{obs}}=0.9$. We also repeated the whole process for different values of $N$ and determined the minimum number of EBBH observations needed to constrain $\mathcal{F}\mathrm{_{GC}}$ to an arbitrarily chosen $\Delta \mathcal{F}\mathrm{_{GC}}= 0.2$ wide $68$ percent (one-sigma) confidence interval (note that for symmetrical errors $\Delta \mathcal{F}\mathrm{_{GC}} = 0.2$ corresponds to a $\pm10$ percent range of the total $[0,1]$ interval).

Fig. \ref{fig:KozaiGNADTest} summarizes our results. We find that the PDFs of $\mathcal{F}\mathrm{_{GC}}$ for fixed values of $\mathcal{F}\mathrm{_{GC}^{obs}}$ peak at $\mathcal{F}\mathrm{_{GC}^{obs}}$, which means that our statistic is unbiased. The widths of the 68 percent and 90 percent intervals decrease with increasing values of $|\mathcal{F}\mathrm{_{GC}^{obs}}-0.5|$, hence the widest intervals correspond to $\mathcal{F}\mathrm{_{GC}^{obs}}=0.5$. Also note that the plots for different values of $\mathcal{F}\mathrm{_{GC}^{obs}}$ are symmetric to $\mathcal{F}\mathrm{_{GC}^{obs}}=0.5$, which means that they would practically be the same if instead of gravitational capture, the secular Kozai-Lidov mechanism was considered. As expected, the 68 percent and 90 percent confidence intervals shrink with increasing numbers of simulated observations. The widths of the 68 percent confidence intervals reach $\Delta \mathcal{F}\mathrm{_{GC}} = 0.2$ at $N=21$ for $\mathcal{F}\mathrm{_{GC}^{obs}}=0.1$ (and for $\mathcal{F}\mathrm{_{GC}^{obs}}=0.9$), at \mbox{$N=43$} for $\mathcal{F}\mathrm{_{GC}^{obs}}=0.3$ (and for $\mathcal{F}\mathrm{_{GC}^{obs}}=0.7$), and at $N=46$ for $\mathcal{F}\mathrm{_{GC}^{obs}}=0.5$. Therefore, we conclude that in the idealized case of having no parameter reconstruction errors, a few tens of EBBH observations are sufficient to constrain $\mathcal{F}\mathrm{_{GC}}$ to a $\Delta \mathcal{F}\mathrm{_{GC}}=0.2$ wide one-sigma confidence interval.

\subsection{Results with non-zero parameter estimation errors}
\label{ssec:KozaiResErr}

In a realistic situation, reconstructed $e_{\mathrm{LIGO}}$ values are liable to estimation errors. We investigated their effect in our statistical test (described in Section \ref{ssec:KozaiRes}) by assuming two different magnitudes for estimation errors on $e_{\mathrm{LIGO}}$ (from now on, denoted by $\Delta e_{\mathrm{LIGO}}$), which we take into account by exchanging each no-error $e_{\mathrm{LIGO}}$ value with a randomized value taken from a Gaussian distribution centered on $e_{\mathrm{LIGO}}$ and with $\sigma = \Delta e_{\mathrm{LIGO}}$ (note that these Gaussian distributions were truncated outside the allowed $e_{\mathrm{LIGO}}\in[0,1]$ range). The two $\Delta e_{\mathrm{LIGO}}$ values we used in our investigations were $\Delta e_{\mathrm{LIGO}}= 10^{-2}$ and $\Delta e_{\mathrm{LIGO}}= 10^{-1}$. Choosing $\Delta e_{\mathrm{LIGO}}= 10^{-2}$ is supported by predictions on plausible errors for observations of EBBHs at $\sim1$ Gpc given in \cite{gondan2018} and \cite{gondankocsis2018}. Thus $\Delta e_{\mathrm{LIGO}}= 10^{-1}$ is an overestimation of the $e_{\mathrm{LIGO}}$ estimation errors realistically achievable in the future.

\begin{figure}
	\includegraphics[width=\columnwidth]{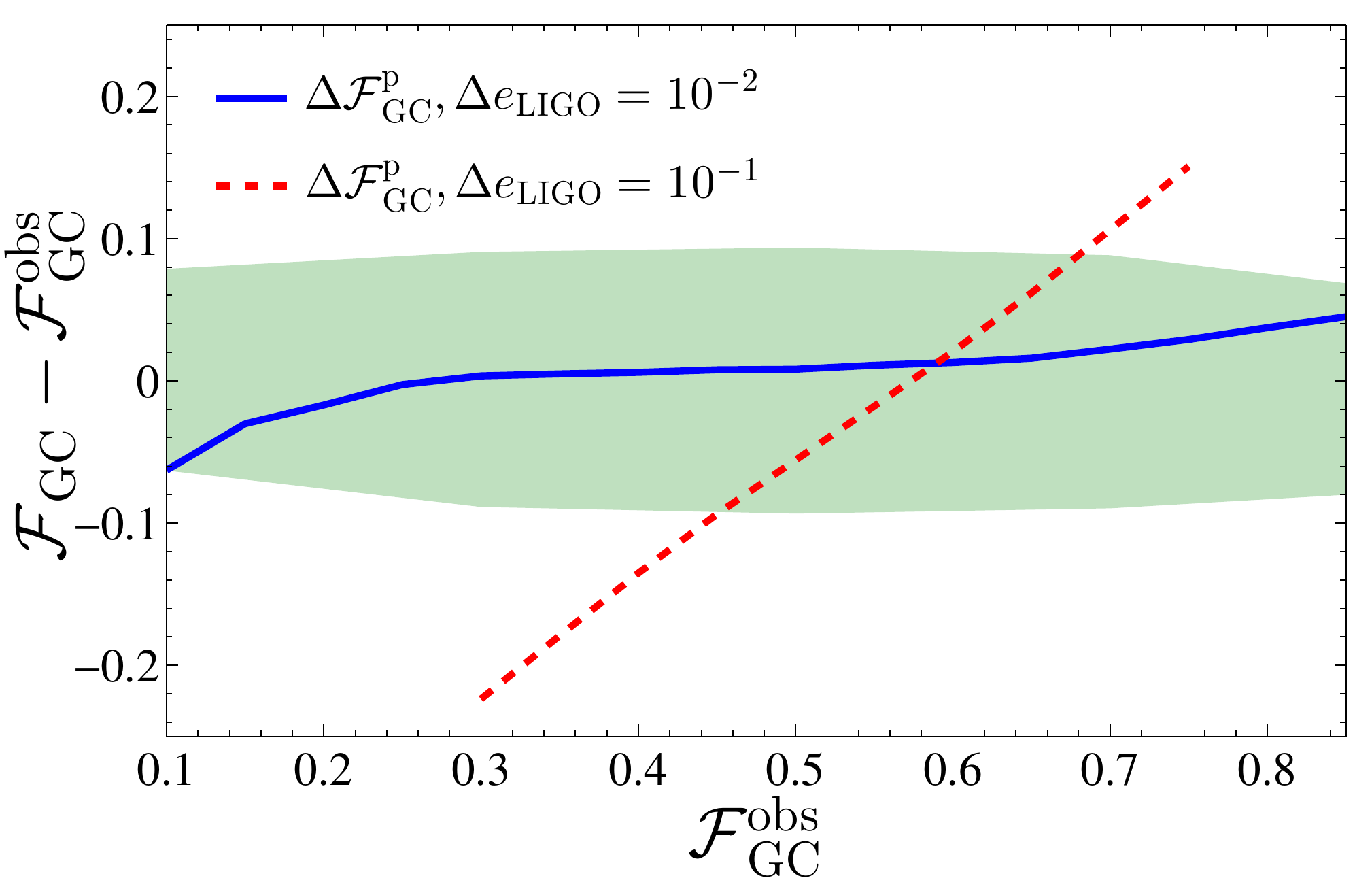}
    \caption{Shift of the peak of the $\mathcal{F}\mathrm{_{GC}}-\mathcal{F}\mathrm{_{GC}^{obs}}$ distribution ($\Delta \mathcal{F}\mathrm{_{GC}^{p}}$) as a function of $\mathcal{F}\mathrm{_{GC}^{obs}}$, for $N = 50$ observed EBBHs, and for $\Delta e_{\mathrm{LIGO}} = 10^{-2}$ (solid blue line) and $\Delta e_{\mathrm{LIGO}} = 10^{-1}$ (dashed red line); see Section \ref{ssec:KozaiResErr} for details. We also show the one-sigma confidence intervals of the distributions of $\mathcal{F}\mathrm{_{GC}}$ obtained in the zero-error ($\Delta e_{\mathrm{LIGO}} = 0$) case (shaded green area). In the case of $\Delta e_{\mathrm{LIGO}} = 10^{-2}$ the shifted peak remains within the one-sigma confidence interval of statistical errors, while for $\Delta e_{\mathrm{LIGO}} = 10^{-1}$ this shift can exceed the interval limits by large amounts. Note that for $\Delta e_{\mathrm{LIGO}} = 10^{-1}$ we only plot $\Delta \mathcal{F}\mathrm{_{GC}^{p}}$ in the valid range when $\mathcal{F}\mathrm{_{GC}}\in [0,1]$.}
    \label{fig:DeltakErr}
\end{figure}

By repeating the analysis described in Section \ref{ssec:KozaiRes} with $N$-element samples of such non-zero error $e_{\mathrm{LIGO}}$ values, we have found that $\mathcal{F}\mathrm{_{GC}^{obs}}$ becomes a biased estimator of $\mathcal{F}\mathrm{_{GC}}$, even when $N$ goes to infinity. This is due to the fact that the $e_{\mathrm{LIGO}}$ values estimated with errors do not follow the zero-error $e_{\mathrm{LIGO}}$ distributions shown in Fig. \ref{fig:eLigo} that we use as bases in our statistical test. For an example case of choosing $N=50$, the peak of the $\mathcal{F}\mathrm{_{GC}}-\mathcal{F}\mathrm{_{GC}^{obs}}$ distribution ($\Delta \mathcal{F}\mathrm{_{GC}^{p}}$) shifts around $(\mathcal{F}\mathrm{_{GC}}-\mathcal{F}\mathrm{_{GC}^{obs}})=0$ by a varying amount that is a function of $\mathcal{F}\mathrm{_{GC}^{obs}}$. Fig. \ref{fig:DeltakErr} shows the amount of this shift as a function of $\mathcal{F}\mathrm{_{GC}^{obs}}$ for both $\Delta e_{\mathrm{LIGO}}= 10^{-2}$ and \mbox{$\Delta e_{\mathrm{LIGO}}= 10^{-1}$}.

We have also found that the widths of the 68 percent confidence intervals of the $\mathcal{F}\mathrm{_{GC}}$ distributions do not change significantly when estimation errors are introduced, but remain essentially the same as in the zero-error case described in Section \ref{ssec:KozaiRes}. Since the dependence of $\Delta \mathcal{F}\mathrm{_{GC}^{p}}$ on $\mathcal{F}\mathrm{_{GC}^{obs}}$ is heavily dependent on $\Delta e_{\mathrm{LIGO}}$, we can only draw conclusions on the magnitude of these shifts for the different $\Delta e_{\mathrm{LIGO}}$ values. For example, we can compare the magnitude of the shifts to the one-sigma confidence intervals of the $\mathcal{F}\mathrm{_{GC}}$ distributions in the zero-error case (shown by a shaded green area in Fig. \ref{fig:DeltakErr}), which closely resembles the corresponding one-sigma confidence intervals for the two non-zero error cases. We have found that with $\Delta e_{\mathrm{LIGO}} = 10^{-1}$ the shift of the peak can reach $\Delta \mathcal{F}\mathrm{_{GC}^{p}} > 0.2$. However with $\Delta e_{\mathrm{LIGO}} = 10^{-2}$, this shift remains within the one-sigma confidence interval of statistical errors, with a maximum shift of $|\Delta \mathcal{F}\mathrm{_{GC}^{p}}| \simeq 0.06$ when $\mathcal{F}\mathrm{_{GC}^{obs}} \in [0.1,0.85]$. Hence, we conclude that in practice $e_{\mathrm{LIGO}}$ estimation errors should be kept at the $\Delta e_{\mathrm{LIGO}} \leq 10^{-2}$ level to be able to constrain the branching ratios between different EBBH formation channels with this method to a $\Delta \mathcal{F}= 0.2$ wide one-sigma confidence interval.

\section{Conclusions}
\label{sec:Conclusion}

We calculated the probability density functions of $e_{\mathrm{LIGO}}$ ($e$ at $f_{\mathrm{GW}}=10$ Hz) of EBBHs formed in GNs through gravitational capture and through the secular Kozai-Lidov mechanism (see Section \ref{ssec:RecPar}), using Monte Carlo simulations described in Section \ref{ssec:MonteCarlo} and \ref{ssec:KozaiNum}. We have found, supplementing the results of \cite{hoang2018}, that $\sim 10$ percent of binary BHs formed through the Kozai-Lidov mechanism in GNs have eccentricities $e_{\mathrm{LIGO}}>0.1$ (see Section \ref{ssec:RecPar}), and that $\sim 75$ percent of binary BHs formed through gravitational capture in GNs have eccentricities $e_{\mathrm{LIGO}}>0.1$. Our simulations of EBBHs formed through gravitational capture extend the simulations of \cite{gondan2017} by introducing randomized masses for the central SMBHs. It is important to note that the $e_{\mathrm{LIGO}}$ distributions presented here may be different if we assume a different underlying model for GNs, for example if we assume that BHs form disks inside GNs as \cite{szolgyen2018} has suggested, or if we assume a different radial dependence of the number density of BHs compared to the one we considered in Section \ref{ssec:Bahcall}. 

Beyond our main results described in the previous paragraph, we have also carried out a statistical test that can be used in the future to constrain the branching ratios (denoted by $\mathcal{F} \in [0,1]$ for a certain formation channel) between the two main EBBH formation channels in GNs with future GW observations of EBBHs (see Section \ref{ssec:KozaiRes} and \ref{ssec:KozaiResErr}). Using these tests we have found that with design aLIGO, a few tens of observations are sufficient for every observed value of $\mathcal{F}\mathrm{^{obs}}$ to constrain $\mathcal{F}$ to a $\Delta \mathcal{F}= 0.2$ wide one-sigma confidence interval (see Section \ref{ssec:KozaiRes} for details). We also supplemented our study with additional tests in which we considered non-zero parameter estimation errors. We have found that $\mathcal{F}\mathrm{^{obs}}$ becomes a biased estimator of $\mathcal{F}$ even if the number of EBBH observations approaches infinity. We conclude that $e_{\mathrm{LIGO}}$ estimation errors should be kept at the feasible $\Delta e_{\mathrm{LIGO}} \leq 10^{-2}$ level \citep{gondan2018,gondankocsis2018} to be able to constrain the branching ratios between the two main channels to a $\Delta \mathcal{F}= 0.2$ wide one-sigma confidence interval (see Section \ref{ssec:KozaiResErr} for details).

The expected detection rate of EBBHs formed through gravitational capture is \mbox{$\mathcal{O}(1-100$ yr$^{-1}$)} with aLIGO detectors operating at design sensitivity \citep{oleary2009}. This aLIGO detection rate may also be higher than \mbox{$\sim$100 yr$^{-1}$} if the BH mass function extends to masses over $25$ $M_{\odot}$ \citep{oleary2009}. Several LIGO and Virgo observations of such high mass BHs have been made recently \citep[e.g.][]{abbott2016,abbott2017a,abbott2017b}. The detection rate of EBBHs formed through the secular Kozai-Lidov mechanism is expected to be of the same order as the detection rate of EBBHs formed through gravitational capture. Based on these detection rates, the minimum observation time needed to constrain the branching ratios between these two channels to a $\Delta \mathcal{F}= 0.2$ wide one-sigma confidence interval is \mbox{$\mathcal{O}(0.1-10$ yr)}, if $e_{\mathrm{LIGO}}$ estimation errors can be kept at the feasible $\Delta e_{\mathrm{LIGO}} \leq 10^{-2}$ level. Note that our tests were carried out while assuming that future observations of EBBHs will be dominated by EBBHs formed in GNs, and thus our results should be treated with due caution.

Note that during our analysis we assumed the search pipeline applied in the future for finding GW signals of EBBHs and reconstructing their parameters is equally sensitive for different eccentricities. This assumption may be idealistic, and must be tested in the future when such pipelines become available. Nevertheless, the selection effect that the limited sensitivity of a specific pipeline introduces can be taken into account as corrections in our predictions for the detected $e_{\mathrm{LIGO}}$ distributions, and thus in our statistical tests as well.

\section*{Acknowledgements}

We would like to thank L\'{a}szl\'{o} Gond\'{a}n and Bence Kocsis for their valuable comments on the manuscript. We would also like to thank L\'{a}szl\'{o} Gond\'{a}n for providing us a set of numerical codes we used in our simulations. J\'{a}nos Tak\'{a}tsy, Bence B\'{e}csy, and P\'{e}ter Raffai were supported by the \'{U}NKP-16-1 (J.T.), \'{U}NKP-17-2 (B.B.), and \'{U}NKP-17-4 (P.R.) New National Excellence Programs of the Ministry of Human Capacities.









\bibliographystyle{mnras}
\balance
\bibliography{GN_EBBHpaper}{}


\bsp	
\label{lastpage}
\end{document}